\documentclass[doc]{apa6}
\usepackage[style=apa6,sortcites=true,sorting=nyt,backend=biber,natbib=true]{biblatex}
\DeclareLanguageMapping{american}{american-apa}
\newtoggle{cleanup}
\toggletrue{cleanup}

\iftoggle{cleanup}{
  \newcommand{\reviewhbm}[1]{{#1}}
}{
  \usepackage{lineno}
  \linenumbers
  \modulolinenumbers[2]

  \newcommand*\patchAmsMathEnvironmentForLineno[1]{
     \expandafter\let\csname old#1\expandafter\endcsname\csname #1\endcsname
     \expandafter\let\csname oldend#1\expandafter\endcsname\csname end#1\endcsname
     \renewenvironment{#1}{\linenomath\csname old#1\endcsname}{\csname oldend#1\endcsname\endlinenomath}}

  \newcommand*\patchBothAmsMathEnvironmentsForLineno[1]{\patchAmsMathEnvironmentForLineno{#1}\patchAmsMathEnvironmentForLineno{#1*}}

  \AtBeginDocument{
  \patchBothAmsMathEnvironmentsForLineno{equation}
  \patchBothAmsMathEnvironmentsForLineno{align}
  \patchBothAmsMathEnvironmentsForLineno{flalign}
  \patchBothAmsMathEnvironmentsForLineno{alignat}
  \patchBothAmsMathEnvironmentsForLineno{gather}
  \patchBothAmsMathEnvironmentsForLineno{multline}

  \newcommand{\reviewhbm}[1]{\textbf{#1}}
  }
}

\newcommand{\new}[1]{#1}
\newcommand{\review}[1]{#1}
\newcommand{\rereview}[1]{{#1}}

\usepackage[T1]{fontenc}
\usepackage[utf8]{inputenc}
\usepackage[english]{babel}
\usepackage{csquotes}
\usepackage{booktabs}
\usepackage{caption}

\usepackage{amssymb}
\usepackage{mathtools}
\usepackage{textcomp}
\usepackage{physics}
\usepackage{bm}
\usepackage{xspace}
\newcommand*{\eg}{e.g.,\@\xspace}
\newcommand*{\ie}{i.e.,\@\xspace}

\usepackage{hyperref}
\hypersetup{
  colorlinks   = true, %
  urlcolor     = blue, %
  linkcolor    = blue,  %
  citecolor    = blue  %
}

\usepackage[export]{adjustbox}
\usepackage{flafter}

\usepackage[section]{placeins}
\usepackage[algoruled]{algorithm2e}
\usepackage[nameinlink,capitalise,english,noabbrev]{cleveref}

\usepackage{floatrow}
\graphicspath{{images/}}
\DeclareGraphicsExtensions{.png,.jpg}

\newcommand{\KL}{\text{KL}}
\newcommand{\Real}{\mathbb{R}}
\newcommand{\bval}[1]{$b$ = #1 s/mm$^2$}
\DeclareMathOperator*{\argmin}{arg\,min}

\journal{Human Brain Mapping}

    \title{Harmonization of diffusion MRI datasets with adaptive dictionary learning}
    \shorttitle{Harmonization of diffusion MRI datasets}

    \author{Samuel St-Jean, Max A. Viergever, Alexander Leemans}
    \authornote{Corresponding author: Samuel St-Jean, email: samuel@isi.uu.nl}

    \affiliation{Image Sciences Institute, University Medical Center Utrecht, Heidelberglaan 100, 3584 CX Utrecht, the Netherlands}

\abstract{%
Diffusion magnetic resonance imaging can indirectly infer the
microstructure of tissues and provide metrics subject to normal variability in a population.
Potentially abnormal values may yield essential information to support analysis of controls and patients cohorts,
but subtle confounds could be mistaken for purely biologically driven variations amongst subjects.
In this work, we propose a new harmonization algorithm based on adaptive dictionary learning to mitigate the unwanted
variability caused by different scanner hardware while preserving the natural biological variability of the data.
Our harmonization algorithm does not require paired training datasets, nor spatial registration or matching spatial resolution.
Overcomplete dictionaries are learned iteratively from all datasets at the same time with an adaptive regularization criterion.
removing variability attributable to the scanners in the process.
The obtained mapping is applied directly in the native space of each subject towards a scanner-space.
The method is evaluated with a public database which consists of two different protocols acquired on three different scanners.
Results show that the effect size of the four studied diffusion metrics is preserved while removing variability attributable to the scanner.
Experiments with alterations using a free water compartment, which is not simulated in the training data,
shows that the modifications applied to the diffusion weighted images are preserved
in the diffusion metrics after harmonization, while still reducing global variability at the same time.
The algorithm could help multicenter studies pooling their data by removing scanner specific confounds, and increase statistical power in the process.}


\begin{document}

\maketitle

\section*{Data availability statement}

The datasets from the CDMRI 2017 challenge are available from the organizers upon request at
\url{https://www.cardiff.ac.uk/cardiff-university-brain-research-imaging-centre/research/projects/cross-scanner-and-cross-protocol-diffusion-MRI-data-harmonisation}.
Code for the harmonization algorithm itself is available at \url{https://github.com/samuelstjean/harmonization}.

\section*{Funding statement}

Samuel St-Jean \reviewhbm{is mainly} supported by the Fonds de recherche du Québec - Nature et technologies (FRQNT) (Dossier 192865)
\reviewhbm{and partly supported by the Natural Sciences and Engineering Research Council of Canada (NSERC)
[funding reference number BP--546283--2020] and the FRQNT (Dossier 290978).}
This research is supported by VIDI Grant 639.072.411 from the Netherlands Organization for Scientific Research (NWO).
The funding agencies were not involved in the design, data collection nor interpretation of this study.

\section*{Conflict of interest statement}

The authors have no conflict of interest to disclose.

\section*{Ethics approval statement}

The original data acquisition was approved by Cardiff University School of Psychology ethics committee.
Written informed consent was obtained from all subjects.

\section*{Permission to reproduce material from other sources}

Some figures in the manuscript have been adapted from open access manuscripts available under the CC-BY license as indicated in their respective legend.

\section*{Acknowledgments}

We would like to thank Chantal Tax for providing us with the evaluation masks and testing datasets from the challenge.
The data were acquired at the UK National Facility for In Vivo MR Imaging of Human Tissue Microstructure located in CUBRIC funded by the EPSRC (grant EP/M029778/1), and The Wolfson Foundation.
Acquisition and processing of the data was supported by a Rubicon grant from the NWO (680-50-1527), a Wellcome Trust Investigator Award (096646/Z/11/Z), and a Wellcome Trust Strategic Award (104943/Z/14/Z).
This database was initiated by the 2017 and 2018 MICCAI Computational Diffusion MRI committees (Chantal Tax, Francesco Grussu, Enrico Kaden, Lipeng Ning, Jelle Veraart, Elisenda Bonet-Carne, and Farshid Sepehrband)
and CUBRIC, Cardiff University (Chantal Tax, Derek Jones, Umesh Rudrapatna, John Evans, Greg Parker, Slawomir Kusmia, Cyril Charron, and David Linden).

\section*{ORCID}

Samuel St-Jean - \url{https://orcid.org/0000-0002-8092-2974}

Alexander Leemans - \url{https://orcid.org/0000-0002-9306-6126}

\setcounter{secnumdepth}{3}

\section{Introduction}
\label{sec:introduction}

Diffusion weighted magnetic resonance imaging (dMRI) is a noninvasive imaging technique that can indirectly infer the
microstructure of tissues based on the displacement of water molecules.
As dMRI only offers an indirect way to study, \eg the brain microstructure, analysis of dMRI datasets includes multiple processing steps to ensure adequate correction
of acquisition artifacts due to subject motion or eddy current induced distortions, amongst others \citep{Tournier2011}.
Quantitative scalar measures of diffusion can be extracted from the acquired datasets, such as the apparent diffusion coefficient (ADC) or fractional anisotropy (FA)
as computed from diffusion tensor imaging (DTI) \citep{Basser1994,Basser1996}, with a plethora of other measures and diffusion models nowadays available \citep{Tournier2019a,Assemlal2011b}.
These measures are subject to normal variability across subjects and potentially  abnormal values or
features extracted from dMRI datasets may yield essential information to support analysis of controls and patients cohorts \citep{Johansen-Berg2009,Jones2011a}.

As small changes in the measured signal are ubiquitous due to differences in scanner hardware \citep{Sakaie2018},
software versions of the scanner or processing tools \citep{Sakaie2018,Gronenschild2012},
field strength of the magnet \citep{Huisman2006} or
reconstruction methods in parallel MRI and accelerated imaging \citep{Dietrich2008,St-jean2018a},
non-negligible effects may translate into small differences in the subsequently computed diffusion metrics.
Subtle confounds affecting dMRI can even be due to measuring at different time points in the cardiac cycle,
leading to changes in the measured values of pseudo-diffusion over the cardiac cycle \citep{Deluca2019a,Federau2013a}.
In the presence of disease, these small variations in the measured signal are entangled in the genuine biological variability,
which is usually the main criterion of interest to discover or analyze subsequently.
This can lead to confounding effects and systematic errors that could be mistaken for purely biologically driven variations amongst subjects.
To mitigate these issues, large-scale studies try to harmonize their acquisition protocols across centers
to further reduce these potential sources of variability \citep{Duchesne2019} or may only use a single scanner without upgrading it for long term studies \citep{Hofman1995a,Hofman2015a}.
The stability brought by keeping the same scanning hardware is however at the cost of potentially missing on improved, more efficient sequences
or faster scanning methods becoming common in MRI \citep{Feinberg2010,Lustig2007,Larkman2001}.
Even by carefully controlling all these sources of variability as much as possible, there still remain
reproducibility issues between scanners of the same model or in scan-rescan studies of dMRI metrics \citep{Vollmar2010,Magnotta2012,Kristo2013a}.
Over the years, many algorithms have been developed to mitigate the variability attributed to non-biological effects in dMRI,
\eg in order to combine datasets from multiple studies and increase statistical power, see \eg \citep{Zhu2019a,Tax2019,Pinto2020} for reviews.
Common approaches consist in harmonizing the dMRI datasets \new{through the coefficients of a spherical harmonics representation}
\citep{Mirzaalian2016a,CetinKarayumak2019,Blumberg2019} or the computed scalar metrics \citep{Fortin2017a,Pohl2016a} to reduce variability between scanners.
Recently, a dMRI benchmark database containing ten training subjects and four test subjects datasets acquired on three scanners with two acquisition protocols
was presented at the computational diffusion MRI (CDMRI) 2017 challenge \citep{Tax2019}.
The publicly available CDMRI database was previously used to compare five harmonization algorithms,
including a previous version of the algorithm we present here, which we use for evaluation.

In this work, we propose a new algorithm based on adaptive dictionary learning to mitigate the unwanted variability caused by different scanner hardware
while preserving the natural biological variability present in the data.
\new{The algorithm is applied directly on the dMRI datasets themselves without using an alternative representation and can be used on datasets acquired at different spatial resolutions or with
a different set of diffusion sensitizing gradients (\ie b-vectors).}
Expanding upon the methodology presented in \citet{St-Jean2016a,St-Jean2017}, overcomplete dictionaries
are learned automatically from the data with an automatic tuning of the regularization parameter \reviewhbm{to balance the fidelity of the reconstruction with sparsity of the coefficients at every iteration.
These dictionaries are either constructed using the data from a given source scanner and used to reconstruct the data from a different target scanner (first set of experiments) or
learned using datasets coming from multiple scanners at once---creating a \enquote{scanner space} in the process (second set of experiments).
One of the improvements of the algorithm is the ability to harmonize datasets acquired with multiple scanners, without explicitly needing to define a source and target scanner as is usually done.
This new formulation also does not need to match the gradient directions (\ie the b-vectors) of the other datasets.}
\rereview{In the first set of experiments,} these dictionaries are used to reconstruct the data \reviewhbm{with a dictionary from a different target} scanner, removing variability present in the source scanner in the process.
Mapping across different spatial resolutions can be obtained by adequate subsampling of the dictionary.
\reviewhbm{In the second set of experiments, the test datasets are altered with simulations mimicking edema while the training datasets are left untouched. We}
show that the harmonization algorithm preserves the natural variability of the data, even if these alterations are not part of the training datasets.
This is done by mapping all the datasets towards a global scanner-space,
which \rereview{can be done for multiple scanners at once \textit{without} paired datasets or spatial registration of subjects} to do so.
\rereview{Removing the prerequisite of} paired datasets for training \rereview{makes the algorithm} easy to apply for hard to
acquire datasets (\eg patients with Alzheimer's, Parkinson's or Huntington's disease) or when pooling datasets from unrelated studies that are acquired in separate centers.
This makes our proposed method readily applicable for pre-existing and ongoing studies that would like to remove variability caused by non-biological or systematic effects in their data analyzes.

\section{Theory}
\label{sec:theory}
\subsection{The dictionary learning algorithm}
\label{sec:algo}

Dictionary learning \citep{Mairal2009b,Elad2006a} aims to find a set of basis elements to efficiently approximate a given set of input vectors.
\reviewhbm{We follow here in general our previous formulation from \citep{Tax2019} which} optimizes both the representation $\bm{D}$ (called the dictionary or the set of atoms) and
the coefficients $\bm{\alpha}$ of that representation (called the sparse codes) as opposed to using a fixed basis (\eg Fourier, wavelets, spherical harmonics).
A dictionary can be chosen to be overcomplete (\ie more column than rows) as the algorithm is designed to only select a few atoms
to approximate the input vector with a penalization on the $\ell_1$-norm of $\bm{\alpha}$ to promote a sparse solution.
Applications in computer vision with the goal to reduce visual artifacts include demosaicking \citep{Mairal2009c}, inpainting \citep{Mairal2009b}
and upsampling \citep{Yang2010,Yang2012} amongst others.

In practice, local windows are used to extract spatial and angular neighborhoods of diffusion weighted images (DWIs) \reviewhbm{inside a brain mask}
to create the set of vectors required for dictionary learning as in \citet{St-Jean2016a}.
This is done by first extracting a small 3D region from a single DWI, which we now refer to as a patch.
To include angular information, a set of patches is taken at the same spatial location across DWIs
in an angular neighborhood (as defined by the angle between their associated b-vector on the sphere).
This considers that patches from different DWIs at the same spatial location, but which are in fact not too far on the sphere,
exhibit self-similarity that can be exploited by dictionary learning.
Once this process is done, every set of patches is concatenated to a single vector $\bm{X}$.
All of these vectors $\bm{X}_n$ are then put in a 2D matrix $\Omega = \{\bm{X}_1, \ldots, \bm{X}_n, \ldots\}$, where $n$ denotes one of the individual set of patches.

Once the set of patches $\Omega$ has been extracted, $\bm{D}$ can be initialized by randomly selecting $N$ vectors from $\Omega$ \citep{Mairal2009b}.
With this initial overcomplete dictionary, a sparse vector $\bm{\alpha}_n$ can be computed for each $\bm{X}_n$
such that $\bm{D}$ is a good approximation to reconstruct $\bm{X}_n$, that is $\bm{X}_n \approx \bm{D\alpha}_n$.
This initial approximation can be refined iteratively by sampling randomly $N$ new vectors $\bm{X}_n \in \Omega$ and updating $\bm{D}$ to better approximate those vectors.
At the next iteration, a new set $\bm{X}_n \in \Omega$ is randomly drawn and $\bm{D}$ is updated to better approximate this new set of vectors.
This iterative process can be written as

\begin{equation}
    \argmin_{\bm{D},{\bm{\alpha}}}\  \frac{1}{N}\sum_{n=1}^N \left(\frac{1}{2}\norm{\bm{X}_n - \bm{D\alpha}_n}_2^2 + \lambda_i\norm{\bm{\alpha}_n}_1\right) \text{ s.t. } \norm{\bm{D}_{.p}}_2^2 = 1
    \label{eq:find_D}
\end{equation}
with $\bm{\alpha}_n \in \Real^{p\times1}$ an array of sparse coefficients and $\bm{D}$ the dictionary where each column is constrained to unit $\ell_2$-norm to prevent degenerated solutions.
$\lambda_i$ is a regularization parameter used at iteration $i$ (which is further detailed in \cref{sec:choose_lambda})
to balance the $\ell_2$-norm promoting data similarity and the $\ell_1$-norm promoting sparsity of the coefficients $\bm{\alpha}_n$.
Iterative updates using \cref{eq:find_D} alternate between refining $\bm{D}$ (and holding $\bm{\alpha}$ fixed) and computing $\bm{\alpha}$ (with $\bm{D}$ held fixed) for the current set of $\bm{X}_n$.
As updating $\bm{\alpha}$ needs an optimization scheme, this can be done independently for each $\bm{\alpha}_n$ using coordinate descent \citep{Friedman2010}.
For updating $\bm{D}$, we use the parameter-free closed form update from \citet[Algorithm 2]{Mairal2009b},
which only requires storing intermediary matrices of the previous iteration using $\bm{\alpha}$ and $\bm{X}_n$ to update $\bm{D}$.
Building dictionaries for the task at hand has been used previously in the context of diffusion MRI for denoising \citep{St-Jean2016a,Gramfort2013}
and compressed sensing \citep{Gramfort2013,Merlet2013a,Schwab2018b} amongst other tasks.
Note that it is also possible to design dictionaries based on products of fixed basis or adding additional constraints such as positivity or spatial consistency
to \cref{eq:find_D}, see \eg \citep{Schwab2018b,Vemuri2019} and references therein for examples pertaining to diffusion MRI.

\subsection{Automatic regularization selection}
\label{sec:choose_lambda}

\cref{eq:find_D} relies on a regularization term $\lambda_i$ which can be different for each set of vectors $\bm{X}_n$ at iteration $i$.
It is, however, common to fix $\lambda_i$ for all $\bm{X}_n$ depending on some heuristics
such as the size of $\bm{X}_n$ \citep{Mairal2009b}, the local noise variance \citep{St-Jean2016a} or through a grid search \citep{Gramfort2013}.
In the present work, \reviewhbm{we instead rely on an automatic tuning criterion since datasets acquired on multiple scanners are subject
to different local noise properties and of various signal-to-noise ratio (SNR) spatially.
In addition, the datasets do not need to be at the same spatial resolution; defining a single scalar value for the regularization parameter
as done in previous works is therefore not straightforward anymore.
In this work,} a search through a sequence of candidates $\{\lambda_0, \ldots, \lambda_s, \ldots, \lambda_\text{last} \}$,
which is automatically determined for each individual $\bm{X}_n$, is instead employed.
\reviewhbm{The optimal value of $\lambda$ is chosen} by minimizing the Akaike information criterion (AIC) \citep{Akaike1974,Zou2007}
\reviewhbm{as in \citep{Tax2019} or additionally by} using either 3-fold cross-validation (CV) and minimizing the mean squared error.
For the AIC, the number of non-zero coefficients in $\bm{\alpha}_n$ provides an unbiased estimate of degrees of freedom for the model \citep{Tibshirani2012a,Zou2007}.
We use the AIC for normally distributed errors in least-squares problems from \citet{Burnham2004}
\begin{equation}
    AIC_{\lambda_i} = \argmin_{\lambda_s}\ m\log\left(\frac{\norm{\bm{X}_n - \bm{D\alpha}_{\lambda_s}}_2^2}{m}\right) + 2 \text{df}(\bm{\alpha}_{\lambda_s})
    \label{eq:aic_lsq}
\end{equation}
with $m$ the number of elements of $\bm{X}_n$.
In practice, this sequence of $\lambda_s$ is chosen automatically on a log scale
starting from $\lambda_0$ (providing the null solution $\bm{\alpha}_{\lambda_0} = 0$) up to $\lambda_\text{last} = \epsilon > 0$ (providing the regular least squares solution) \citep{Friedman2010}.
The solution $\bm{\alpha}_n$ at $\lambda_{s}$ is then used as a starting estimate for the next value of $\lambda_{s+1}$.
The process can be terminated early if the cost function \cref{eq:find_D} does not change much (\eg the difference between the solution at $\lambda_s$ and $\lambda_{s+1}$ is below $10^{-5}$)
for decreasing values of $\lambda_s$, preventing computation of similar solutions.

\section{Methods}
\label{sec:method}

\reviewhbm{In this section, we detail how a dictionary can be learned to create an implicit mapping between scanners.
This is done by first constructing a target dictionary with datasets acquired on at least one or multiple scanners.
After this target dictionary is constructed, a set of coefficients using the data from a given source scanner is computed,
keeping the precomputed target dictionary fixed during the process.
The resulting reconstructed datasets have implicit features specifically captured by the initial target scanner,
without reconstructing the features only found in the source scanner used to acquire the data initially.}

\subsection{Building an optimal representation across scanners}
\label{sec:rebuild_D}

For harmonization based on dictionary learning, all 3D patches of small spatial and angular local neighborhoods inside a brain mask
were extracted from the available training datasets for a given scanner as done in \citep{St-Jean2016a,Tax2019}.
Since different patch sizes are used depending on the reconstruction task, \cref{sec:recon_challenge,sec:recon} detail each case that we study in this manuscript.
Only patches present inside a brain mask were used for computation and reconstruction.
These patches were reorganized as column arrays $\Omega = \{\bm{X}_1, \ldots, \bm{X}_n, \ldots\}$ with each $\bm{X}_n \in \Real^{m\times1}$ represented as vectors of size $m$.
Each volume was mean subtracted and each patch $\bm{X}_n$ was scaled to have unit variance \citep{Friedman2010}.
Subsequently, features were automatically created from the target scanner datasets using dictionary learning as detailed in \cref{sec:algo}.
A dictionary $\bm{D} \in \Real^{m \times p}$ was initialized with $p$ vectors $\bm{X}_{m\times1} \in \Omega$ randomly chosen,
where $\bm{D}$ is set to have twice as many columns as rows (\ie $p = 2m$) \review{as previously done in \citet{St-Jean2016a,St-Jean2017}}.
Updates using \cref{eq:find_D} were carried for 500 iterations using a batchsize of $N=32$.
The coefficients $\bm{\alpha}_n$ were unscaled afterwards.

Once a dictionary $\bm{D}$ has been computed, the new, harmonized representation (possibly from a \emph{different} scanner)
can be obtained by computing $\bm{\alpha}_n$ for every $\bm{X}_n \in \Omega$.
As $\bm{D}$ was created to reconstruct data from a chosen target scanner, it contains generic features tailored to this specific target scanner
that are not necessarily present in the set of patches $\Omega$ extracted from a different scanner.
As such, reconstruction using $\bm{D}_\text{target}$ created from $\Omega_\text{target}$
can be used to map $\Omega_\text{source}$ towards $\Omega_\text{target}$, that is
$\bm{X}_{n_\text{harmonized}} = \bm{D_\text{target} \alpha}_n$ by using $\bm{X}_{n_\text{source}}$ and holding $\bm{D}_\text{target}$ fixed while solving \cref{eq:find_D} for $\bm{\alpha}_n$.
These specially designed features from $\Omega_\text{target}$ are not necessarily present in $\Omega_\text{source}$,
therefore eliminating the source scanner specific effects, as they are not contained in $\bm{D}_\text{target}$.

\begin{figure}
    \includegraphics[width=\linewidth]{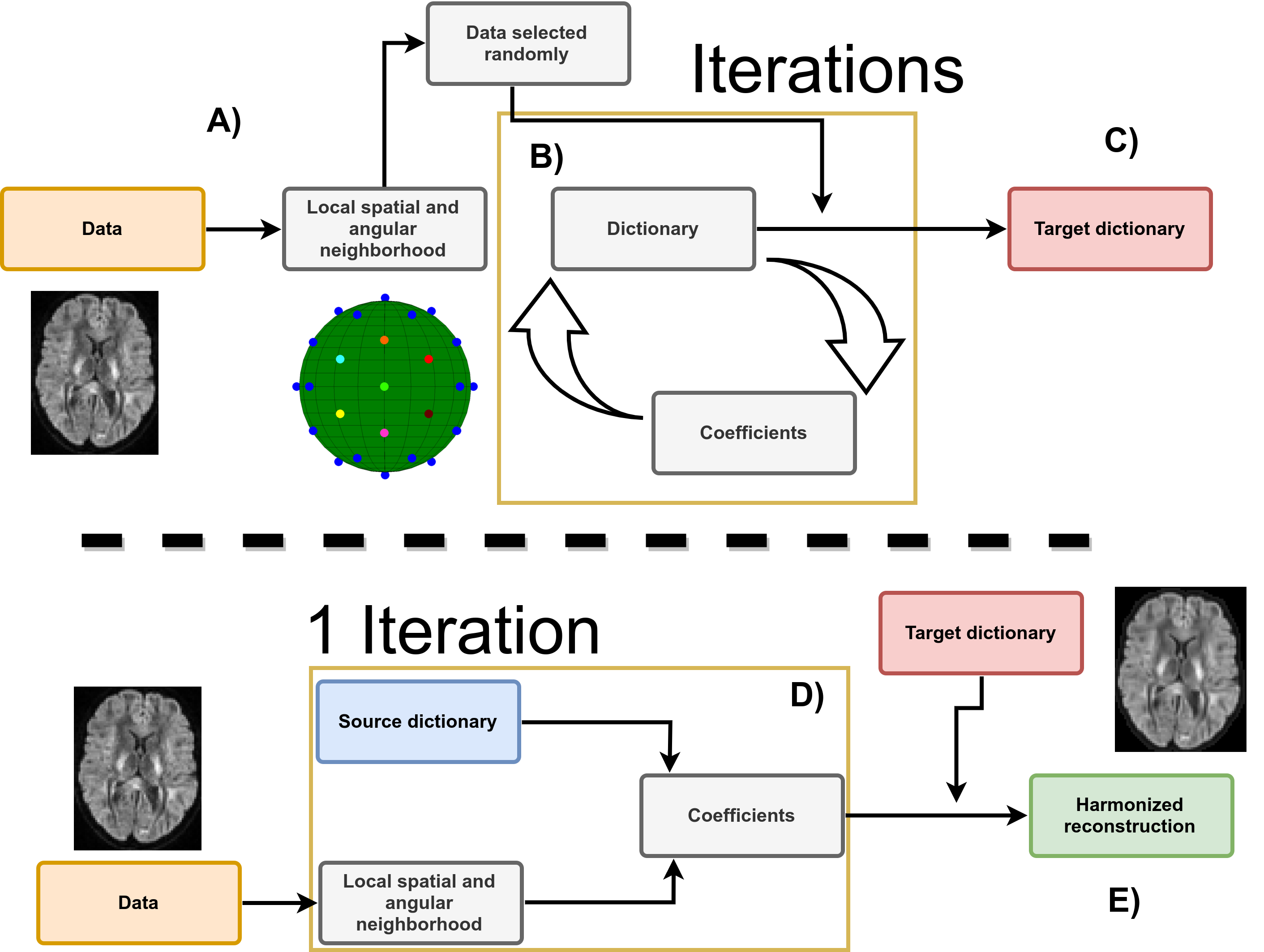}
    \caption{Schematic representation of the harmonization between scanners with adaptive dictionary learning.
    \textbf{A)} Local patches are decomposed into vectors $\bm{X}_n$ and a random subset is used to initialize the dictionary $\bm{D}$.
        \reviewhbm{For harmonization to a scanner-agnostic space, $\bm{D}$ is initialized with data drawn from all scanners.}
    \textbf{B)} A new set of patches is drawn at every iteration and the dictionary is refined iteratively
        by alternating updates for the coefficients $\bm{\alpha}$ and the dictionary $\bm{D}$ using \cref{eq:find_D}.
    \textbf{C)} After a set number of iterations, this target dictionary $\bm{D}$ can now be used to reconstruct data from a potentially different dataset.
    \textbf{D)} A set of coefficients is computed for each patch $\bm{X}_n$ of the input dataset with a source dictionary.
        For harmonization tasks, the source \reviewhbm{dictionary is the target dictionary obtained from a different scanner in} step \textbf{C)} \reviewhbm{and of the same size}.
        For upsampling tasks, the source dictionary is a downsampled version of the target dictionary.
        \reviewhbm{When $\bm{D}$ is constructed from datasets acquired on multiple scanners, the reconstruction step
        removes variability intrinsic to a given dataset which is not present in the remaining scanners as $\bm{D}$ would not have captured this variability.}
    \textbf{E)} The harmonized reconstruction for each patch $\bm{X}_n$ is obtained by multiplying the target dictionary $\bm{D}$ and the coefficients $\bm{\alpha}_n$.
    }
    \label{fig:diagram}
\end{figure}

Downsampling $\bm{D}_\text{target}$ into $\bm{D}_\text{small}$ can also be used to reconstruct data
at a different resolution than initially acquired by creating an implicit mapping between two different spatial resolutions.
This is done by finding the coefficients $\bm{\alpha}$ by holding $\bm{D}_\text{small}$ fixed when solving \cref{eq:find_D},
but using $\bm{D}_\text{target}$ for the final reconstruction such that $\bm{X}_{n_\text{harmonized}} = \bm{D}_\text{target}\bm{\alpha}_n$.
This reconstruction with the full sized dictionary provides an upsampled version of $\bm{X}_n$,
the implicit mapping being guaranteed by sharing the same coefficients $\bm{\alpha}_n$ for both reconstructions.
A similar idea has been exploited previously for the 3D reconstruction of T1w images by \citet{Rueda2012a} and in diffusion MRI by \citet{St-Jean2017} in the context of single image upsampling.
The general reconstruction process for the harmonization of datasets between scanners is illustrated in \cref{fig:diagram}.
Our implementation of the harmonization algorithm is detailed in \cref{sec:appendix} and also available in both source form
and as a Docker container at \url{https://github.com/samuelstjean/harmonization} \citep{St-Jean2019e}.

\subsection{Reconstruction tasks of the challenge}
\label{sec:recon_challenge}

For the reconstruction in task 1 (matched resolution scanner-to-scanner mapping), the dictionary $\bm{D}_\text{target}$ was created
using patches of size $3\times 3\times 3$ with 5 angular neighbors and one \review{randomly chosen} \bval{0} image in each block.
\review{We chose these parameters as they have previously been shown to offer a good trade-off between accuracy
and computation time in a previously published denoising task \citep{St-Jean2016a}.
The angular patch size (\ie how many DWIs are included across gradients) is chosen to include all volumes at the same angular distance on the sphere as in \citet{St-Jean2016a}.}
Optimization for constructing $\bm{D}_\text{target}$ with \cref{eq:find_D} was performed using 3-fold CV and
reconstruction of the final harmonized datasets was done with either CV or minimizing the AIC with \cref{eq:aic_lsq} in two separate experiments.
The datasets from the GE scanner were reconstructed using the dictionary built from the Prisma or Connectom scanner datasets for their respective harmonization task.
For the reconstruction in task 2 (spatial and angular resolution enhancement), patches of different spatial sizes were extracted
from the images at higher resolution (patches of size $5\times 5\times 5$ for the Prisma scanner and $6\times 6\times 6$ for the Connectom scanner)
and used for the dictionary learning algorithm as described in \cref{sec:algo}.
\review{In this task, the target datasets patch size was chosen so that the ratio between the original patch size of $3\times 3\times 3$
and the target patch size matches the ratio in spatial resolution between the harmonized datasets as previously done in \citet{St-Jean2017}.}
Under the hypothesis that a larger patch is a good representation for its lower resolution counterpart when downsampled,
each column of the optimized dictionary $\bm{D}_\text{target}$ was resized to a spatial dimension of $3\times 3\times 3$
and the coefficients $\bm{\alpha}$ computed for this lower resolution dictionary $\bm{D}_\text{small}$.
The patches were finally reconstructed by multiplying the original dictionary $\bm{D}_\text{target}$ with the coefficients $\bm{\alpha}$.
This creates a set of upsampled patches from the GE scanner that are both harmonized and at the same spatial resolution as either the Prisma or the Connectom datasets.
All reconstruction tasks were computed overnight on our computing server using 100 cores running at 2.1 GHz.
On a standard desktop with a 4 cores 3.5 GHz processor, rebuilding one dataset took approximately two hours and 30 minutes with the AIC criterion.

\subsection{Evaluation framework of the challenge}
\label{sec:eval_challenge}

The original challenge requested the participants to match the original gradient directions of the source to the target datasets
and evaluated various scalar metrics on the diffusion weighted images.
In our original submission, this matching was done with the truncated spherical harmonics (SH) basis
of order~6 \citep{Descoteaux2007b} on the source dataset and sampling the basis at the gradient directions from the target scanner.
In the present manuscript, we chose instead to evaluate the metrics directly in the original gradient directions as they are rotationally invariant,
saving one interpolation step in the process as it could potentially introduce unwanted blurring of the data.
The metrics used in the original evaluation were the apparent diffusion coefficient (ADC) and the fractional anisotropy (FA) from diffusion tensor imaging (DTI)
and the rotationally invariant spherical harmonic (RISH) features of order 0 (RISH~0) and order 2 (RISH~2) of the SH basis, see \citet{Tax2019} for additional details.
As our evaluation framework is slightly different, \review{a direct numerical comparison with the results previously reported in the CDMRI challenge
is not possible, even if we use the same metrics, as the exact way to compute the metrics was not made available to the participants.
This unfortunately prevents us from replicating exactly the challenge or to exclude poorly performing regions
as was done in the original evaluation, making a comparison between the previously reported results impossible.}
We compare our new approach \reviewhbm{using automatic regularization with both the AIC and CV criterion}
against our initial version of the harmonization algorithm \reviewhbm{(which included interpolation of the DWIs using the SH basis)}
and a baseline reference prediction created by trilinear interpolation from the source to the target scanner in the spirit of the original challenge.

\subsection{Datasets and experiments}
\label{sec:experiments}

\begin{figure}
    \includegraphics[width=\textwidth]{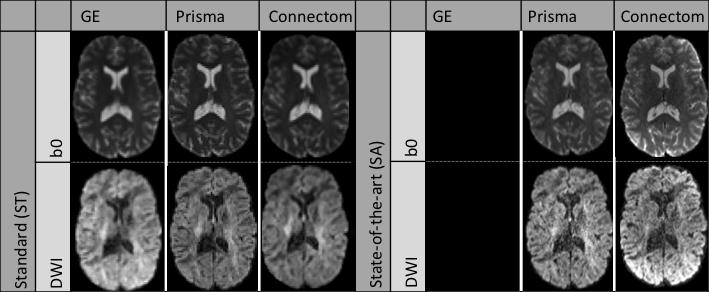}
    \caption{Example \bval{0} images (top row) and \bval{1200} images (bottom row) for a single subject acquired on the three scanners after preprocessing.
    The standard protocol (ST) is shown on the left and the state-of-the-art protocol (SA) is shown on the right.
    Note that the challenge asked participants to harmonize the GE ST protocol towards the two other scanners, but no SA protocol is available for the GE scanner.
    The figure is adapted from \citet{Tax2019}, available under the CC-BY 4.0 license.}
    \label{fig:datasets}
\end{figure}

We used the datasets from the \review{CDMRI} 2017 harmonization challenge \citep{Tax2019},
consisting of ten training subjects and four test subjects acquired on three different scanners (GE, Siemens Prisma and Siemens Connectom)
using different gradient strength (40 mT/m, 80 mT/m and 300 mT/m, respectively) with two acquisition protocols.
\rereview{The study was originally approved by Cardiff University School of Psychology ethics committee and written informed consent was obtained from all subjects.}
Experiments are only reported for the four test subjects, which are later on denoted as subjects 'H', 'L', 'M' and 'N'.
The standard protocol (ST) consists of 30 DWIs acquired at 2.4 mm isotropic with a b-value of \bval{1200},
3 \bval{0} images for the GE datasets, 4 \bval{0} images for the Siemens datasets and TE = 98 ms.
Note that the TR is cardiac gated for the GE datasets while the Siemens datasets both use TR = 7200 ms.
The state-of-the-art (SA) protocol for the Siemens scanners contains 60 DWIs with a b-value of \bval{1200} and 5 \bval{0} images.
The Prisma datasets were acquired with a spatial resolution of 1.5 mm isotropic and TE / TR = 80 ms / 4500 ms.
The Connectom datasets were acquired with a spatial resolution of 1.2 mm isotropic and TE / TR = 68 ms / 5400 ms.
Most of the acquisition parameters were shared for the SA protocol which are
listed in \cref{tab:protocol} with full details of the acquisition available in \citet{Tax2019}.
Standard preprocessing \review{applied by the challenge organizers on the datasets}
includes motion correction, EPI distortions corrections and image registration for each subject across scanners.
The SA protocols were additionally corrected for gradient nonlinearity distortions.
These datasets are available upon request from the organizers at \url{https://www.cardiff.ac.uk/cardiff-university-brain-research-imaging-centre/research/projects/cross-scanner-and-cross-protocol-diffusion-MRI-data-harmonisation}.
\cref{fig:datasets} shows an example of the acquired datasets for a single subject.

\begin{table}
\resizebox{\textwidth}{!}{%
\begin{tabular}{@{}llllll@{}}
    \toprule
    Scanner                            & GE 40 mT/m             & \multicolumn{2}{c}{Siemens Prisma 80 mT/m}               & \multicolumn{2}{c}{Siemens Connectom 300 mT/m}         \\
    \cmidrule(r){1-1}
    \cmidrule(lr){2-2}
    \cmidrule(lr){3-4}
    \cmidrule(lr){5-6}
    Protocol                           & Standard (ST)          & Standard (ST)          & State-of-the-art (SA)          & Standard (ST)          & State-of-the-art (SA)          \\
    \midrule
    Sequence                           & TRSE                   & PGSE                   & PGSE                           & PGSE                   & PGSE                           \\
    \# directions per b-value          & 30                     & 30                     & 60                             & 30                     & 60                             \\
    TE {[}ms{]}                        & 89                     & 89                     & 80                             & 89                     & 68                             \\
    TR {[}ms{]}                        & Cardiac gated          & 7200                   & 4500                           & 7200                   & 5400                           \\
    $\Delta/\delta$ {[}ms{]}           &                        & 41.4/26.0              & 38.3/19.5                      & 41.8/28.5              & 31.1/8.5                       \\
    $\delta_1 = \delta_4/\delta_2 = \delta_3$   {[}ms{]}        & 11.23/17.84  &         &                                &                        &                                \\
    Acquired voxel size {[}mm$^3${]}   & 2.4 x 2.4 x 2.4        & 2.4 x 2.4 x 2.4        & 1.5 x 1.5 x 1.5                & 2.4 x 2.4 x 2.4        & 1.2 x 1.2 x 1.2                \\
 Reconstructed voxel size {[}mm$^3${]} & 1.8 x 1.8 x 2.4        & 1.8 x 1.8 x 2.4        & 1.5 x 1.5 x 1.5                & 1.8 x 1.8 x 2.4        & 1.2 x 1.2 x 1.2                \\
    SMS factor                         & 1                      & 1                      & 3                              & 1                      & 2                              \\
    Parallel imaging                   & ASSET 2                & GRAPPA 2               & GRAPPA 2                       & GRAPPA 2               & GRAPPA 2                       \\
    Bandwidth {[}Hz/Px{]}              & 3906                   & 2004                   & 1476                           & 2004                   & 1544                           \\
    Partial Fourier                    & 5/6                    & ---                    & 6/8                            & 6/8                    & 6/8                            \\
    Coil combine                       &                        & Adaptive combine       & Sum of Squares                 & Adaptive combine       & Adaptive combine               \\
    Head coil                          & 8 channel              & 32 channel             & 32 channel                     & 32 channel             & 32 channel                     \\
    \bottomrule
\end{tabular}%
}
\caption{Acquisition parameters of the datasets for the three different scanners.
TRSE: twice-refocused spin-echo, PGSE: pulsed-gradient spin-echo,
\reviewhbm{TE: echo time, TR: repetition time, SMS: Simultaneous multi-slice, Hz/Px: Hertz/Pixel}.
The table is adapted from \citet{Tax2019}, available under the CC-BY 4.0 license.}
\label{tab:protocol}
\end{table}

\subsection{Simulations beyond the challenge}
\label{sec:recon}

To further make our proposed harmonization algorithm widely applicable,
we designed additional experiments beyond the challenge to harmonize data towards a new scanner-space.
As the \review{CDMRI} challenge focused on harmonization of datasets from a source scanner to a target scanner,
the organizers essentially provided matching datasets of all subjects across all scanners.
This data collection would be appropriate, for example, in a longitudinal study design with scanner hardware upgrades during the study and subsequent data analysis.
However, such an experimental setup might not be available in practice when harmonizing datasets from multiple centers or studies
where data collection is done only once per subject \eg to reduce costs associated with scan time or reduce traveling of the participants.

The additional experiments create a new harmonization space \reviewhbm{by randomly sampling datasets from the three scanners} at once \review{to build the target dictionary}
instead of matching the GE datasets to \reviewhbm{a particular target} scanner as in the previous experiments.
To ensure that the scanner effects are properly removed, the test datasets were \review{additionally} altered in a small region with a simulated free water compartment
as described in \cref{sec:fwe_datasets}, \review{creating additional test datasets contaminated with simulated edema}.
\review{These newly created datasets} were never \enquote{seen} by the harmonization algorithm, \review{making it possible to} quantify
if the induced effects are properly reconstructed \review{without discarding the natural biological variability of the datasets},
as \review{these alterations} were not present in the training set in the first place.
This experiment is similar to creating a common space on a larger set of healthy subjects
and finally harmonizing data from the remaining healthy subjects and \enquote{patients} towards this common space.
In our current setup, the harmonization algorithm is not aware that the datasets are in fact from matched subjects and, by design,
could also be used on unpaired training datasets.

\subsection{Alterations of the original datasets}
\label{sec:fwe_datasets}

To create the altered version of the test datasets, a region of 3000 voxels ($15 \times 20 \times 10$ voxels) in the right hemisphere was selected
at the same spatial location in image space.
\reviewhbm{The size of the region is kept constant throughout experiments and subjects to facilitate statistical analysis and comparisons.}
Every voxel in the selected region was separately affected by a free water compartment to mimic infiltration of edema according to
\begin{equation}
    S_{b_\text{altered}} = S_b + f S_0 \exp(-b D_\text{csf})
\end{equation}
with $S_{b_\text{altered}}$ the new signal in the voxel, $S_b$ the original signal in the voxel at b-value $b$ and $S_0$ the signal in the \bval{0} image,
$f$ is the fraction of the free water compartment, which is drawn randomly for every voxel from a uniform distribution $U(0.7, 0.9)$
and $D_\text{csf} = 3 \times 10^{-3} \text{ mm}^2$/s is the nominal value of diffusivity for free water (\eg cerebrospinal fluid (CSF)) at 37 \textcelsius\ \citep{Pasternak2009a,Pierpaoli2004a}.
\reviewhbm{As the individual subjects across scanners are only registered to their counterpart across scanners, the affected region will be approximately (up to errors due to registration)
at the same spatial location in each subject. This location will however be slightly different between subjects, which}
introduces normal variability in terms of the number of white matter and gray matter voxels that would be affected by edema
and their location in a \reviewhbm{cohort of} patients.

\subsection{Evaluation metrics}
\label{sec:eval}
\paragraph{Error and accuracy of predicted metrics}

We reproduced parts of the analyses conducted in the original CDMRI challenge from \citet{Tax2019}, namely the
per voxel error for each metric as computed by the mean normalized error (MNE) and the voxelwise error.
Denoting the target data to be reproduced as \textit{acquired} (Prisma or Connectom scanners) and the source data to be harmonized as \textit{predicted} (GE scanner),
the MNE is defined as \text{MNE = |(predicted - acquired)| / acquired} and the error is defined as \text{error = predicted - acquired}.
\reviewhbm{The MNE is a relative metric, penalizing more when the error is large relative to the target value,
while the error itself only measures the magnitude of the mistake, but can indicate global under or overestimation with the sign of the metric.
A small error with a large MNE would likely indicate that most of the mistakes committed by an algorithm are in regions where the metric of interest is low.}
The original challenge reports values taken either globally in a brain mask, in FreeSurfer regions of interest (ROI) and excluding poorly performing regions
or the median value computed in sliding windows.
Since the masks of these ROIs were not released for the challenge,
we instead report boxplots of the two metrics using the brain masks from the challenge
as this reports the global median error in addition to the global mean error and additional quantiles of their distribution.
To prevent outliers from affecting the boxplots (particularly located at the edges of the brain masks),
we clip the MNE and error values at the lowest 0.1\% and largest 99.9\% for each dataset separately.

\paragraph{Kullback-Leibler divergence as a measure of similarity}

As the voxelwise difference may not be fully indicative of the global trend of the harmonization procedure between datasets (\eg due to registration errors),
we also computed the Kullback-Leibler (KL) divergence \citep{Kullback1951} between the distributions of each harmonized dataset from the GE scanner
and its counterpart from the target scanner for each of the four metrics.
The KL divergence is a measure of similarity between
two probability distributions $P(x)$ and $Q(x)$ where lower values indicate a higher similarity and $\KL(P, Q) = 0$ when $P(x) = Q(x)$.
In its discrete form, the Kullback-Leibler divergence is given by
\begin{equation}
    \KL(P, Q) = \sum_k P_k \log \left(\frac{P_k}{Q_k}\right),
    \label{eq:kldiv}
\end{equation}
where $P_k$ is the \enquote{candidate} probability distribution, $Q_k$ the true probability distribution
and $k$ represents the number of discrete histogram bins.
The measure is not symmetric, that is $\KL(P, Q) \neq \KL(Q, P)$ in general.
We instead use the symmetric version of the KL divergence as originally defined by \citet{Kullback1951}
\begin{equation}
    \KL_{sym} = \KL(P, Q) + \KL(Q, P).
    \label{eq:kldiv_symmetric}
\end{equation}
In practice, a discrete distribution can be constructed from a set of samples by binning and counting the data.
By normalizing each bin so that their sum is 1, we obtain a (discrete) probability mass function.
For each metric, the discrete distribution was created with $k=100$ equally spaced bins.
We also remove all elements with probability 0 from either $P_k$ or $Q_k$ (if any)
to prevent division by 0 in \cref{eq:kldiv}.
\reviewhbm{As the binning procedure does not share the same bins between scanners,
the results can not be compared directly between the Connectom and Prisma scanners.}

\paragraph{Statistical testing and effect size in the presence of alteration}

\review{To evaluate quantitatively if the harmonization algorithm did not remove signal attributable to genuine biological variability, we computed the percentage difference
between the harmonized test datasets in the affected region of 3000 voxels before alteration and after alteration as given by \cref{eq:percentage_diff}

\begin{equation}
    100 \times \abs{\left(\frac{\text{harmonized} - \text{baseline}}{\text{baseline}}\right) - \left(\frac{\text{harmonized\_altered} - \text{baseline\_altered}}{\text{baseline\_altered}}\right)},
    \label{eq:percentage_diff}
\end{equation}

where baseline (resp. harmonized) denotes the datasets before (resp. after) harmonization and \reviewhbm{the suffix altered} indicates the datasets altered with simulated edema.
A value close to 0 therefore indicates that the harmonization procedure performed similarly in reducing variability attributable to differences
in the scanner for harmonization of the regular datasets and in the presence of alteration.
To investigate the magnitude of these differences, we conducted Student's t-test for paired samples for each subject separately \citep{Student1908a}.}
This was done on both the normal datasets (testing between scanners) and the altered datasets (testing between scanners and additionally between the normal and altered datasets).
The p-values from the tests were subsequently corrected for the false discovery rate (FDR) at a level of $\alpha = 0.05$ \citep{Benjamini1995}.
In addition, we also report the effect size of those paired t-tests as computed by Hedges' $g$ \citep{Hedges1981,Lakens2013a}, which we redefine as
\begin{equation}
    g = \frac{\abs{\mu_1 - \mu_2}}{(\sigma_1 + \sigma_2) / 2} \times \left(1 - \frac{3}{4(n_1 + n_2) - 9}\right),
    \label{eq:effsize}
\end{equation}
where $\mu_i$, $\sigma_i$ and $n_i$ are the mean, the standard deviation, and the size of sample $i$, respectively.
A value of $g=1$ indicates that the difference between the means is of one standard deviation,
with larger values indicating larger effect sizes as reported by the difference in the group means.
In the original definition of \citet{Hedges1981}, $g$ is not enforced to be positive.
We instead report the absolute value of $g$ as we do not know \textit{a priori} which mean is larger than the other,
but are only interested in the magnitude of the effect rather than its sign.
With this definition, values of $g$ reported for the test between a given subject for two different scanners which are lower than the reference method
indicate an improvement by removing scanner specific effects.
On the other hand, similar values of $g$ between the reference and the harmonized dataset for a given subject and its altered counterpart on the same scanner
indicates preservation of the simulated effects as it is the \textit{only} difference between these two datasets by construction.

\section{Results}
\label{sec:results}
\subsection{Results from the challenge}
\label{sec:results_challenge}
\paragraph{Mapping between scanners for matched acquisition protocols}

\begin{figure}
    \includegraphics[width=\linewidth]{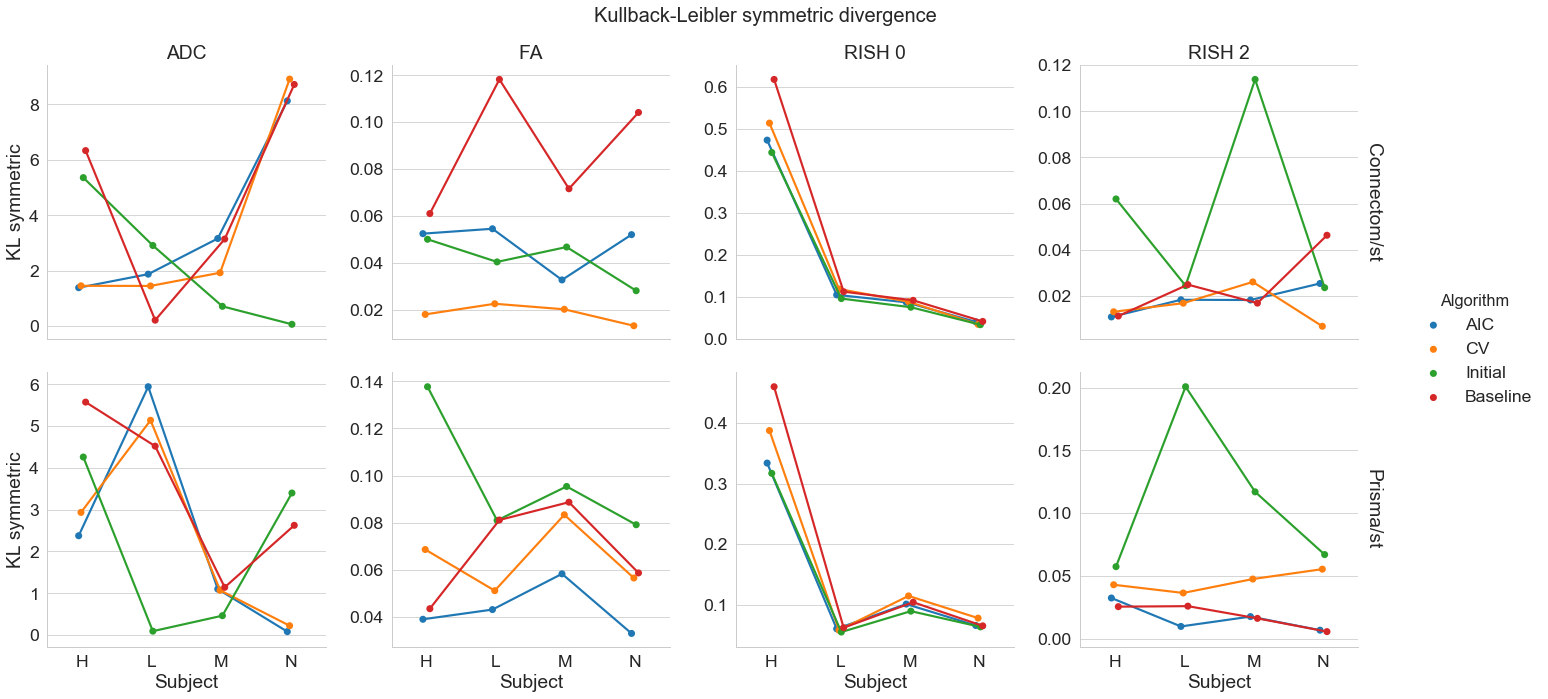}
    \caption{KL symmetric divergence (where lower is better) for the harmonization task at the same resolution
    between the GE ST datasets and the Connectom ST (top row) or the Prisma ST (bottom row) datasets
    on the four test subjects ('H', 'L', 'M' and 'N').
    Each metric is organized by column (ADC, FA, RISH~0 and RISH~2) for the four compared algorithms
    (AIC in blue, CV in orange, our initial version of the harmonization algorithm in green and the baseline comparison in red).}
    \label{fig:st_kl}
\end{figure}

\cref{fig:st_kl} shows the KL symmetric divergence as presented in \cref{sec:eval} for the standard protocol.
In general, the baseline has a higher KL value than the other methods on the Connectom scanner.
The CV based method is generally tied or outperforms the AIC based method.
For the Prisma scanner, results show that the AIC performs best with the CV based method following the baseline reference.
In the case of the ADC metric, our initial algorithm outperforms the three other methods for some subjects.

\begin{figure}
    \includegraphics[width=\linewidth]{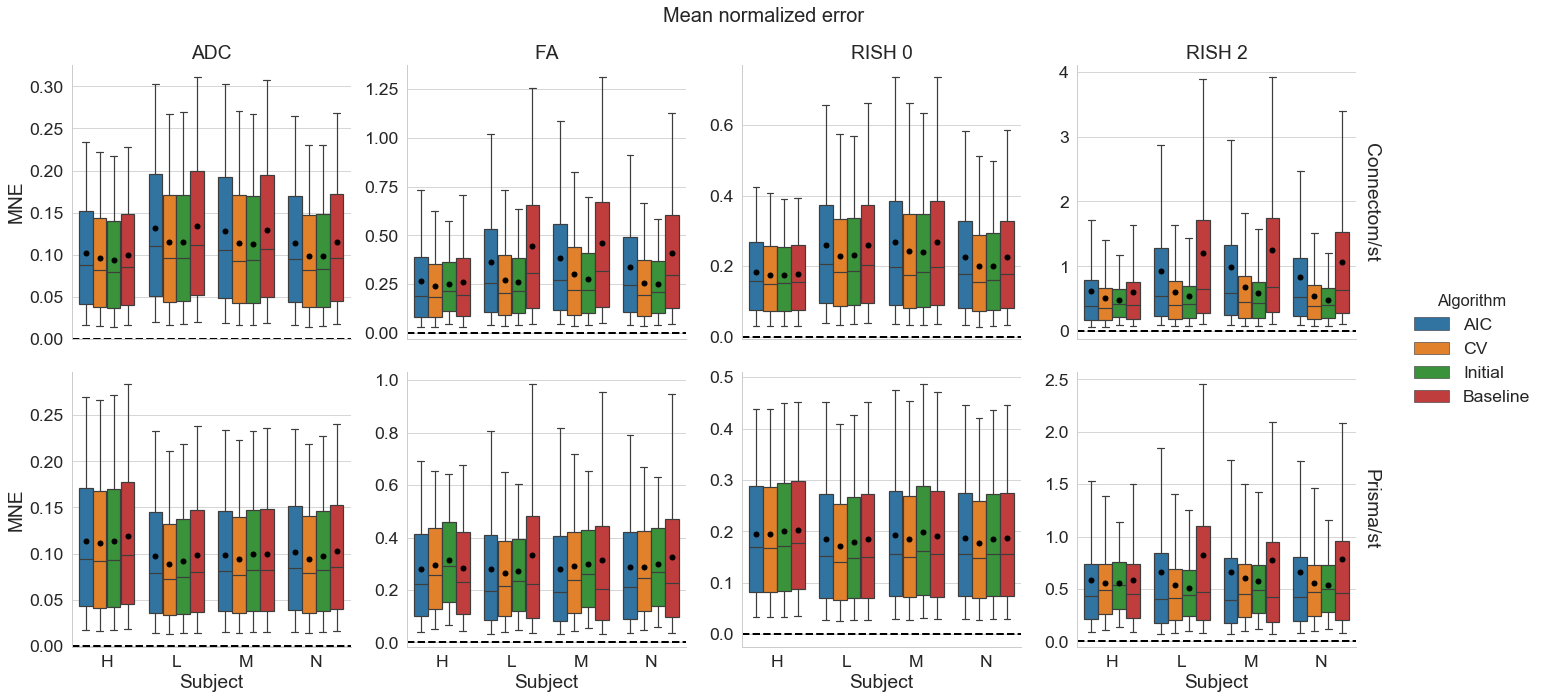}
    \includegraphics[width=\linewidth]{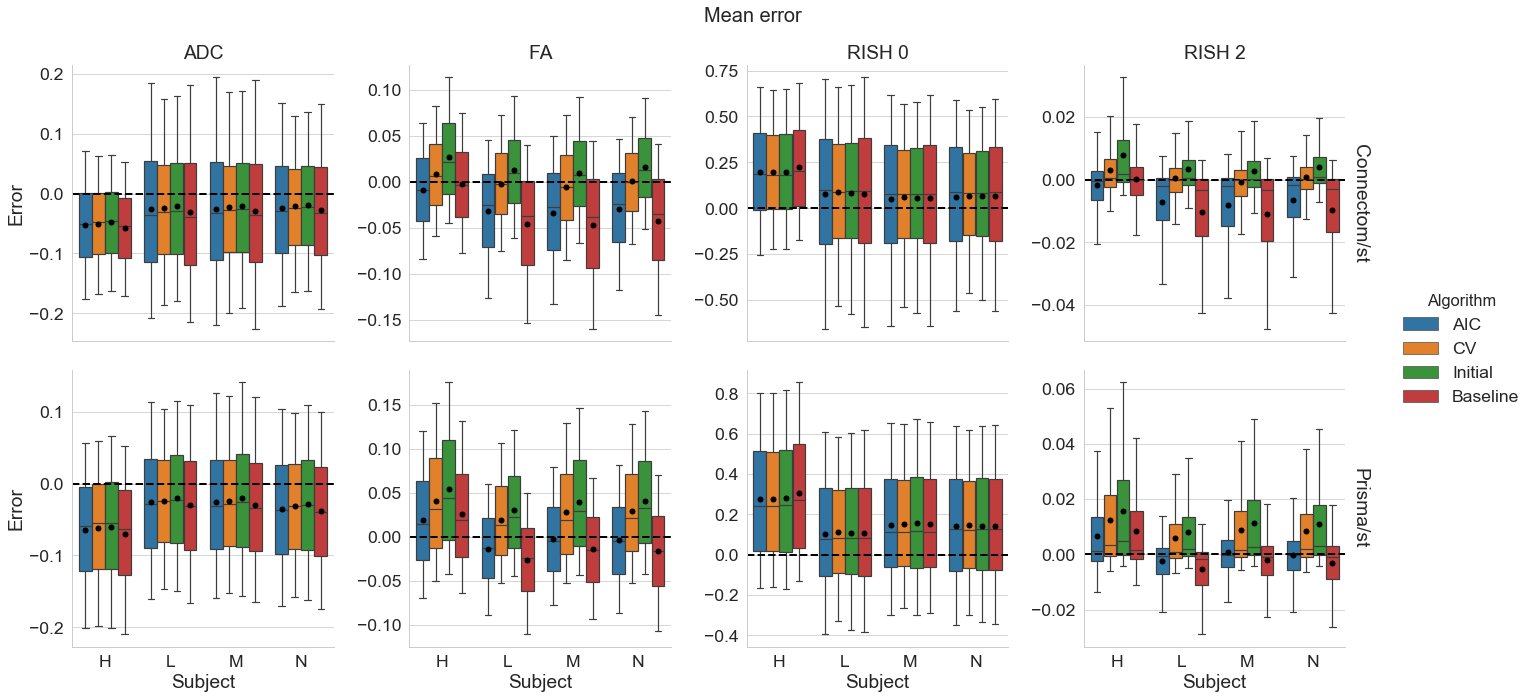}
    \caption{Boxplots of the voxelwise mean normalized error (top) and error (bottom) for each metric,
    following the same conventions detailed in \cref{fig:st_kl}.
    The black dot shows the mean error and the dashed line indicates an error of 0,
    representing a perfect match between the harmonized GE dataset and the dataset for the target scanner.}
    \label{fig:st_error}
\end{figure}

\cref{fig:st_error} shows the distribution (as boxplots) in the absolute mean normalized error and mean error of the four metrics for the standard protocol.
The MNE is almost tied or slightly higher for the baseline method than the alternatives for both scanners.
For the FA and RISH~2 metrics, the baseline error is tied or larger than the other methods.
For the voxelwise error, all methods underestimate the ADC and overestimate the RISH~0 on average
while the FA and RISH~2 metrics show a different pattern depending on the scanner.
For the Connectom scanner, the CV based method generally has an average error around 0 for the FA
while the AIC and our initial algorithm generally overestimate the metric.
The baseline is on the other spectrum and generally underestimates the FA.
On the Prisma scanner, the effect is reversed; there is a general overestimation of the FA
while the error committed by the AIC based method is in general close to 0.
The RISH~2 error follows the same pattern as the FA error on both scanners for the four compared methods.

\paragraph{Mapping between scanners across spatial resolutions}

\begin{figure}
    \includegraphics[width=\linewidth]{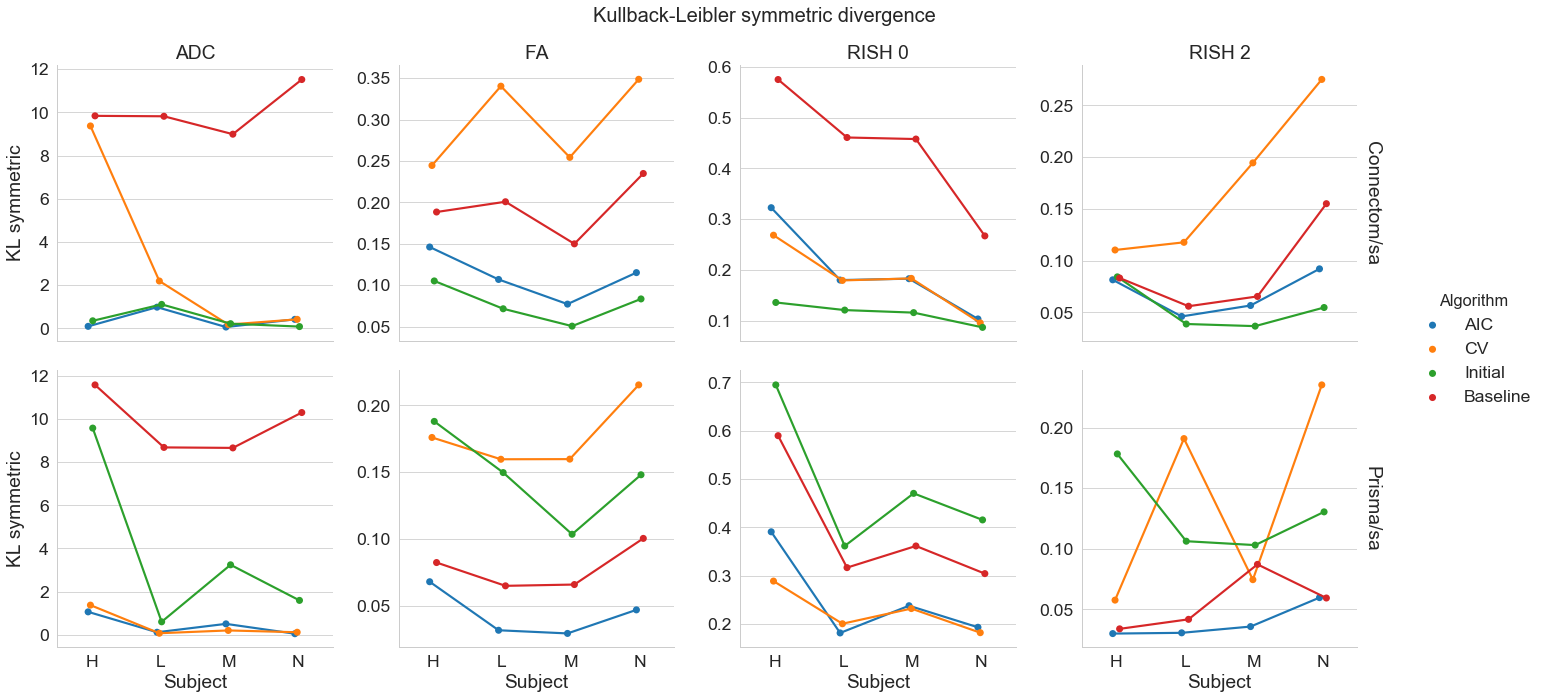}
    \caption{Symmetric KL divergence (where lower is better) for the harmonization task across resolution
    between the GE ST datasets and the Connectom SA (top row) or the Prisma SA (bottom row) datasets.
    The organization is the same as previously used in \cref{fig:st_kl}.}
    \label{fig:sa_kl}
\end{figure}

\new{
\cref{fig:sa_kl} shows the KL symmetric divergence for the second task of the challenge,
mapping the GE ST protocol datasets to the SA protocols of the Prisma or Connectom scanners.
For the Connectom scanner, the AIC based algorithm and our initial algorithm, which is also AIC based, performs best in most cases.
The CV based algorithm also outperforms the baseline method for the ADC and RISH~0 metrics.
For the Prisma scanner, the AIC outperforms most of the compared methods or is tied with the CV.
Notably, the baseline ranks second for the FA and RISH~2 metrics,
but is the worst performer for the ADC and the RISH~0 metrics.
}

\begin{figure}
    \includegraphics[width=\linewidth]{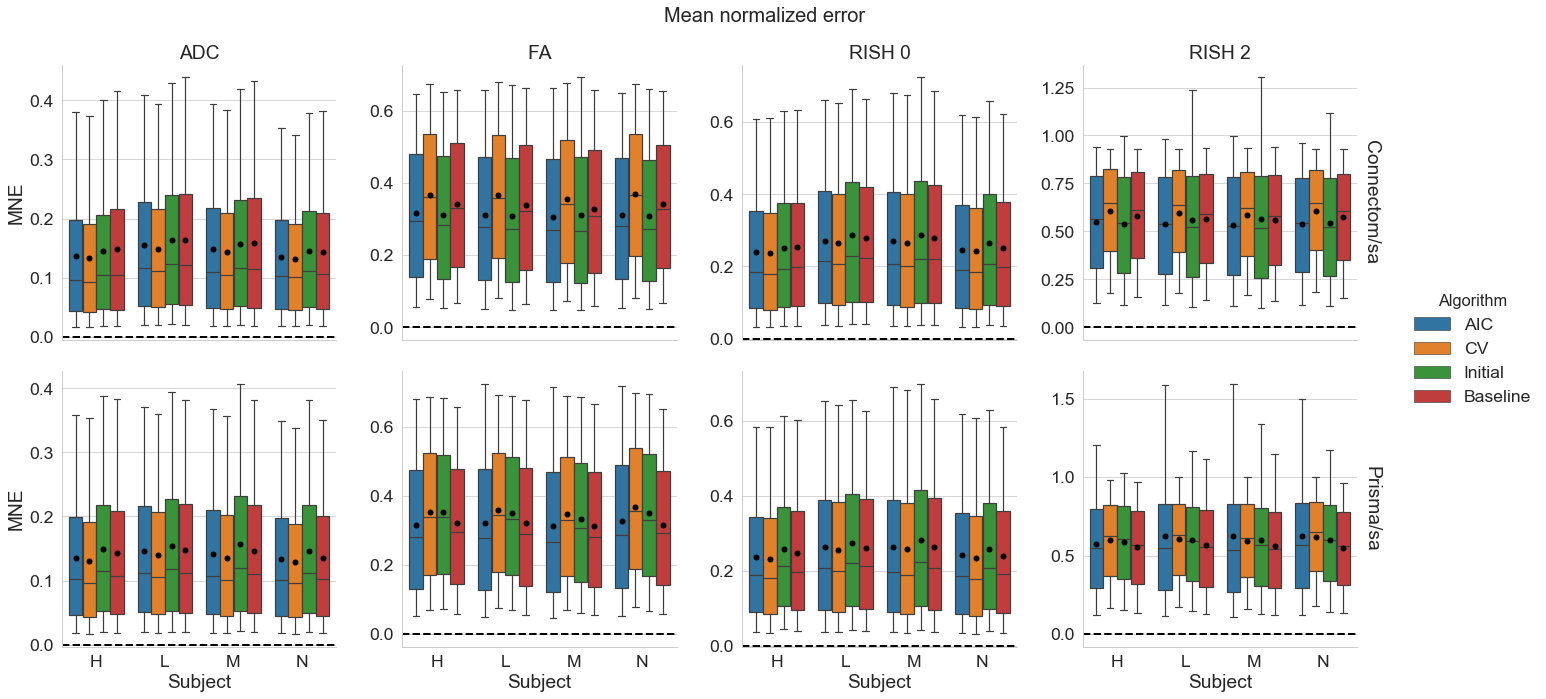}
    \includegraphics[width=\linewidth]{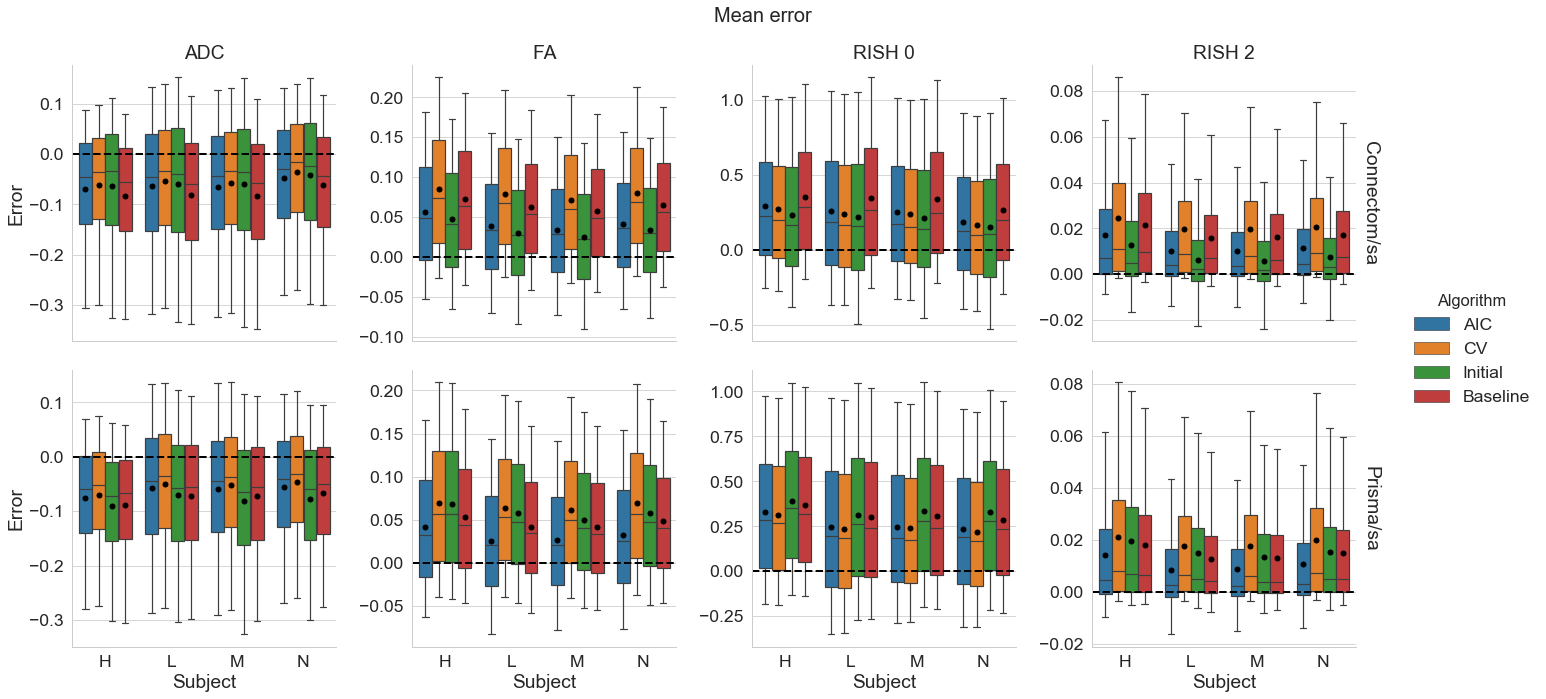}
    \caption{Boxplots of the voxelwise mean normalized error (top) and error (bottom) of each metric for the four algorithms.
    The black dot shows the mean error and the dashed line indicates an error of 0.
    The organization follows the conventions of \cref{fig:st_error}.}
    \label{fig:sa_error}
\end{figure}

\new{
\cref{fig:sa_error} shows results for the absolute mean normalized error and mean error for all algorithms on harmonizing the SA protocol.
For the Connectom scanner, the baseline ranks last for most subjects on the isotropy metrics (ADC and RISH~0)
while it only performs slightly better than the CV based algorithm for the anisotropy metrics (FA and RISH~2).
On the Prisma scanner, results are similar for the ADC and RISH~0 metrics.
For the FA metrics, the best performance is obtained with the AIC based method while
the baseline is better for harmonizing the RISH~2 metric for three of the subjects.

Now looking at the mean error, results show that the ADC metric is underestimated for all methods and on both scanners
with the three methods usually outperforming the baseline comparison.
The FA, RISH~0 and RISH~2 metrics are instead overestimated.
For the FA metric, the AIC and our initial algorithm commit less error on average than the baseline on the Connectom scanner.
On the Prisma scanner, only the AIC has an average error lower than the baseline.
All methods perform better or almost equal on average to the baseline comparison for the RISH~0 metric.
The RISH~2 metric shows a scanner dependent pattern; on the Connectom scanner, the best performing method
is our initial algorithm followed by the AIC based algorithm while on the Prisma scanner, the lowest error is achieved by the AIC based method.
}

\new{
In general, results show that the isotropy metrics (ADC and RISH~0) are subject to global scanner effects
while the anisotropy metrics (FA and RISH~2) may be subject to orientation dependent effects.
These effects are also likely different for each scanner since the gradient strength and timings are different, even if the b-values are matched.
In these experiments, the target scanner is untouched and therefore still contains its own scanner effect
when computing the voxelwise error of each harmonization algorithm.
}

\subsection{Mapping original and altered datasets towards a common space}
\label{sec:results_fwe}

\begin{figure}
    \includegraphics[width=\linewidth]{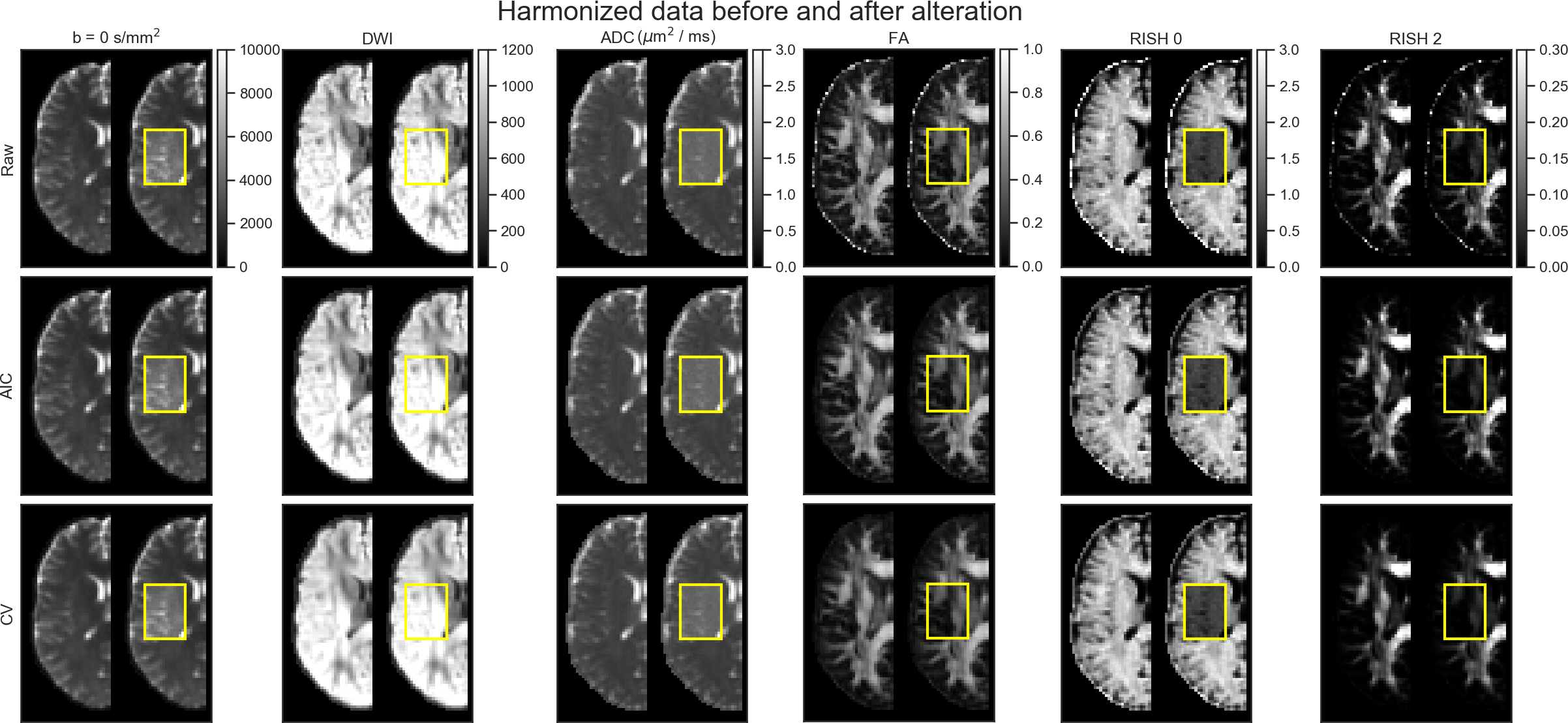}
    \caption{Examplar slice of subject 'H' on the GE scanner as original (left half) and altered (right half) metrics.
    Only the affected portion of the data (\reviewhbm{as shown in the} yellow box) is analyzed in paired statistical testing
    against the same location in the original dataset.
    Each column shows (from left to right) a \bval{0} image, a DWI at \bval{1200}, the FA, ADC, RISH~0 and RISH~2 metrics with a common colorbar per column.
    The top row shows the raw data, the middle row shows the data harmonized using the AIC and the bottom row shows the harmonized data using the CV.
    The \bval{0} image, the DWI and the ADC map increase after adding the free water compartment
    while the FA, RISH~0 and RISH~2 metrics are instead lower in their altered counterpart.}
    \label{fig:datagrid}
\end{figure}

\begin{figure}
    \includegraphics[width=\linewidth]{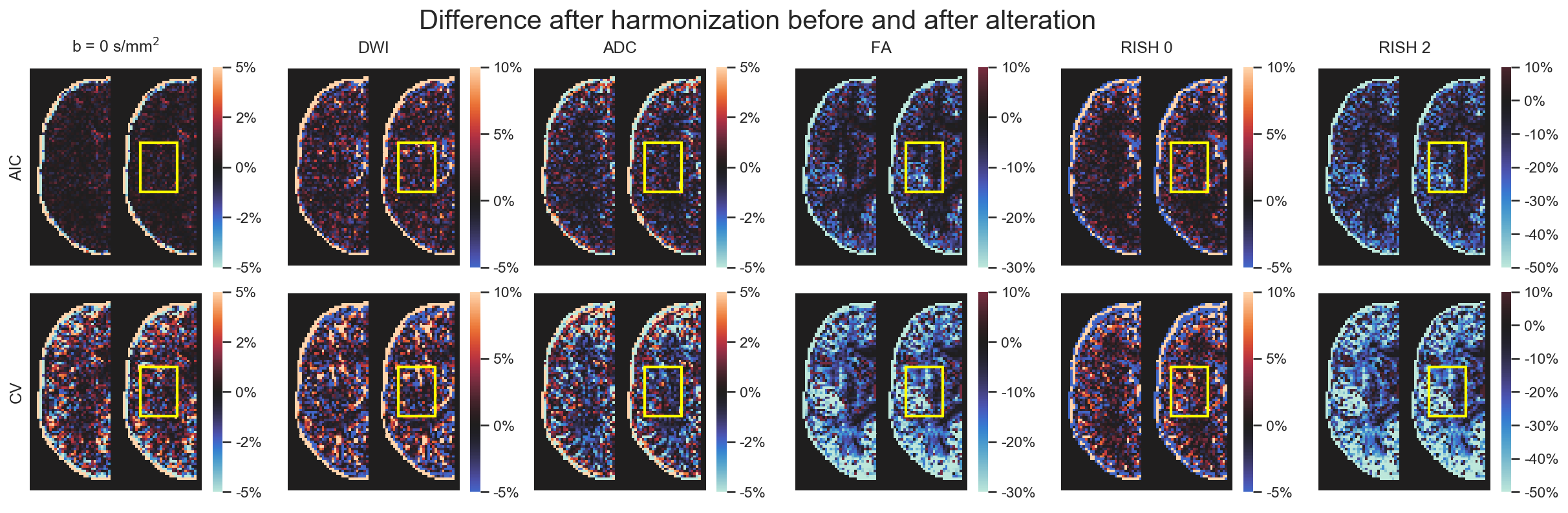}
    \caption{Examplar slice of subject 'H' on the GE scanner as original (left half) and altered (right half) metrics
    \reviewhbm{with the yellow box indicating the altered region specifically}.
    Each column shows (from left to right) a \bval{0} image, a DWI at \bval{1200}, the ADC, FA, RISH~0 and RISH~2 metrics
    with a common colorbar per column as in \cref{fig:datagrid}.
    The top row (resp. the bottom row) shows the relative percentage difference between the harmonized data using the AIC (resp. the CV) and the raw data.
    \reviewhbm{If the affected region is similar in both images, this means that the harmonization algorithm did not remove
    the artificial alterations that were introduced and only removed variability attributable to the scanner equally in both cases.}
    }
    \label{fig:datagrid_diff}
\end{figure}

In these experiments, alterations were made to the test set as previously described in \cref{sec:fwe_datasets}.
As these altered datasets were never used for training, we can quantify the removal of scanner effects and the preservation of the alterations
by comparing solely the altered regions with their original counterpart in each subject, free of processing effects.
In these experiments, the baseline comparison is to not process the datasets at all since the datasets are altered versions of themselves,
therefore not requiring any interpolation or resampling.
As these experiments are outside of the challenge's scope, they are not covered by our initial algorithm
and therefore the \review{\enquote{initial}} category is not presented in this section.
\cref{fig:datagrid} shows the original and altered metrics for one subject on the raw data and after harmonization with the AIC and CV based algorithms
and \cref{fig:datagrid_diff} shows the relative percentage difference between the raw datasets and their harmonized counterpart.
We define the relative percentage difference as $\text{difference} = 100 \times \text{(harmonized - raw) / raw}$.
The alterations are mostly visible on the \bval{0} image while the \bval{1200} image is only slightly affected due to the high diffusivity of the CSF compartment.
However, the differences are visible on the diffusion derived maps, seen as an increase in ADC and a reduction for the FA, RISH~0 and RISH~2 metrics.
Visually, harmonized datasets do not seem different from their original counterpart, but the difference maps show that small differences are present
with the CV method generally showing larger differences than the AIC method.
Notably, the anisotropy metrics (FA and RISH~2) are lower after harmonization while the difference for the isotropy metric (ADC and RISH~0) is distributed around 0.

\begin{figure}
    \includegraphics[width=\linewidth]{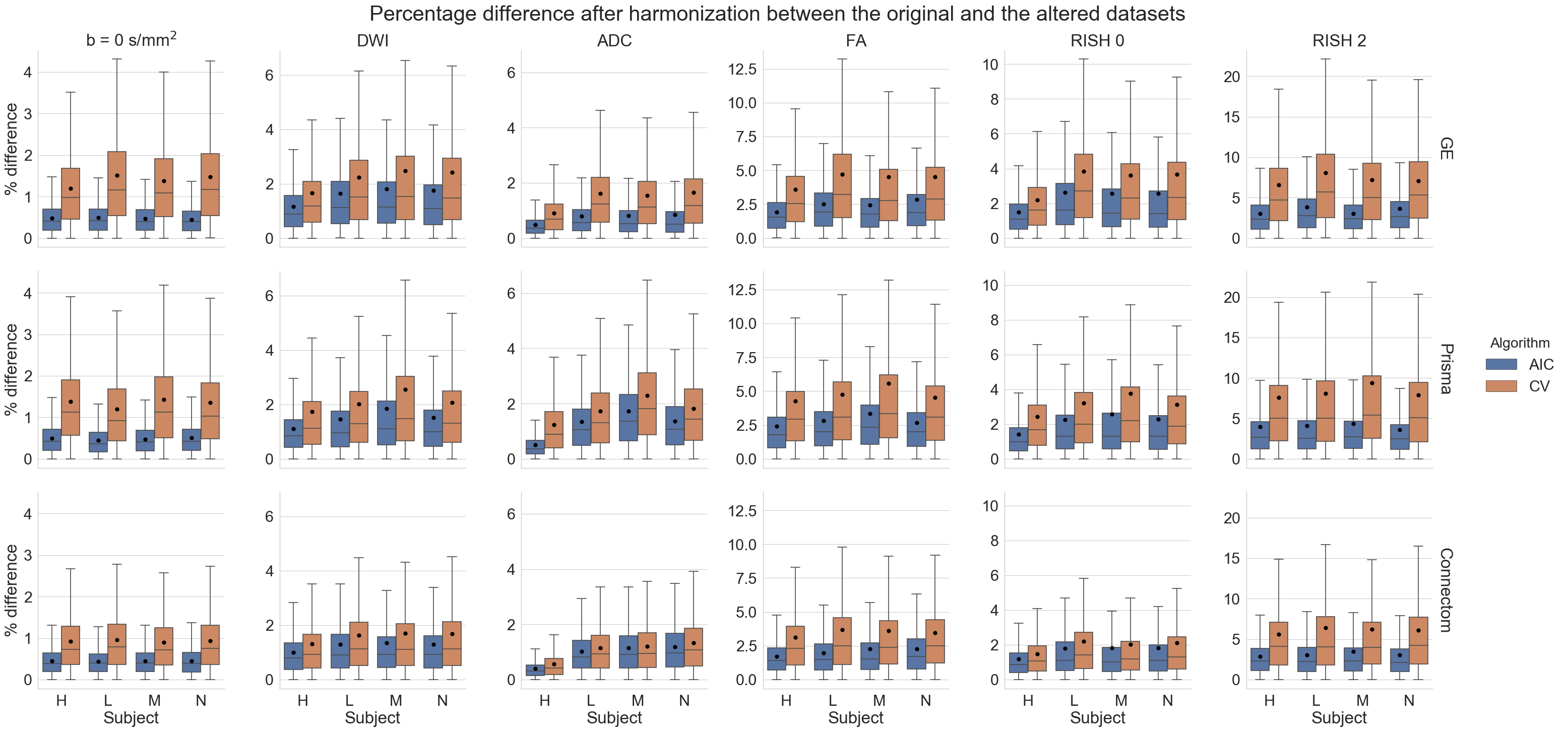}
    \caption{\review{Boxplots of the percentage difference between the harmonized datasets with and without alteration for all subjects for the AIC and CV criterion
    in the altered region only.
    The top row shows the difference for the GE scanner, the middle row for the Prisma scanner and the bottom row for the Connectom scanner.
    A value close to 0 indicates that the harmonization procedure removed a similar amount of the signal in the reference datasets and in the altered datasets.
    }}
    \label{fig:percentage_diff}
\end{figure}

\begin{figure}
    \includegraphics[width=\linewidth]{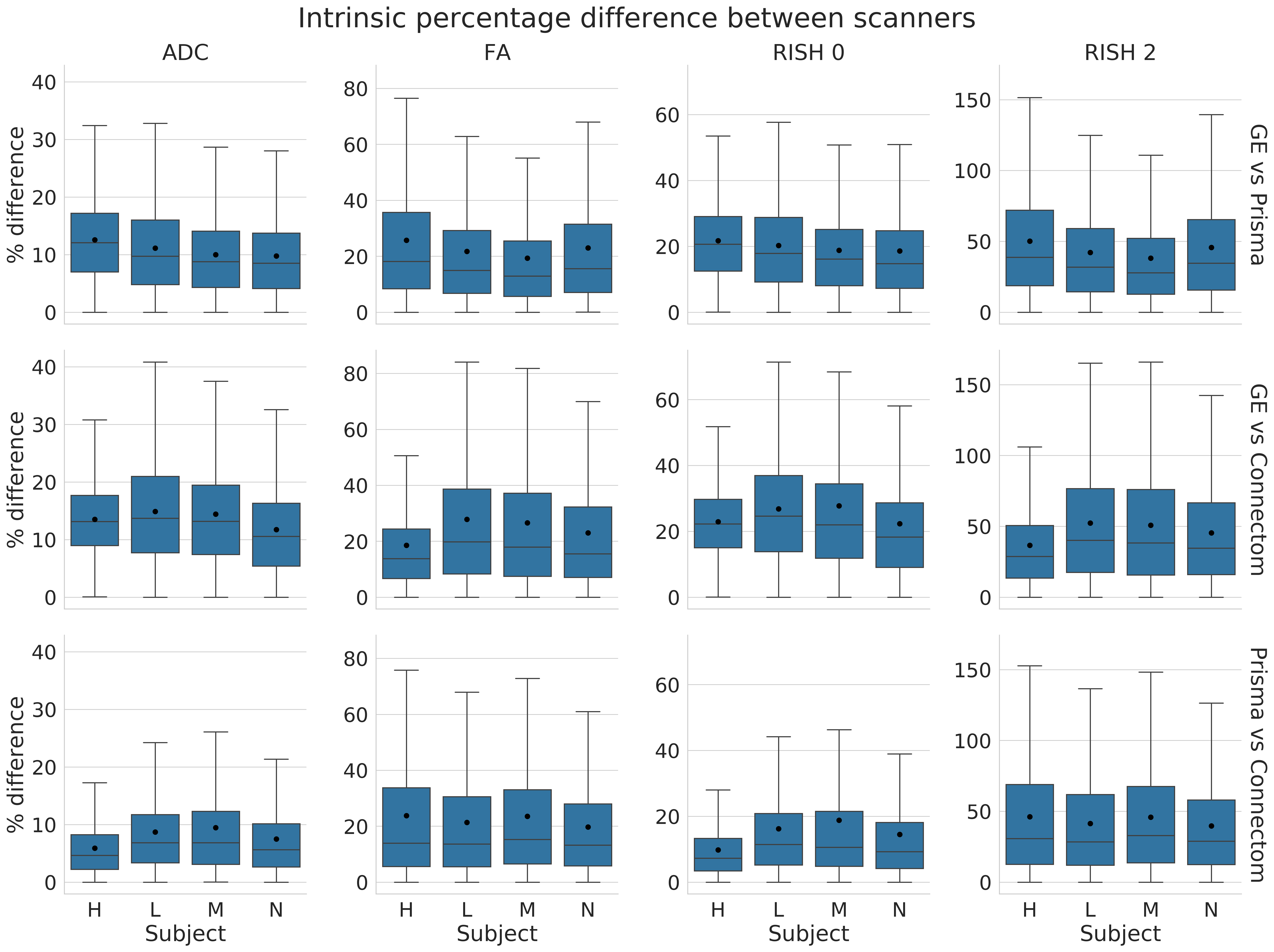}
    \caption{\reviewhbm{Boxplots of the percentage difference (before harmonization) between datasets acquired on different scanners in the selected region before alterations.
    The top row shows the difference between the GE scanner and Prisma scanner, the middle row between the GE and Connectom scanner
    and the bottom row between the Prisma and the Connectom scanner. In this case, the percentage difference is
    computed as $100 \times \frac{\abs{\text{scanner1 - scanner2}}}{\text{(scanner1 + scanner2)} / 2}$, similarly to \cref{eq:percentage_diff}.}}
    \label{fig:percentage_diff_raw}
\end{figure}

\review{\cref{fig:percentage_diff} shows the relative percentage difference as boxplots for all test subjects and all scanners between the altered and normal regions.
A low difference indicates that the signal removed after harmonization is the same in the baseline and altered datasets, that is the algorithm performs
similarly in the presence (or not) of the simulated edema.
The CV algorithm produces larger relative differences than the AIC based algorithm after harmonization between the reference and altered datasets.
The larger differences are in the anisotropy metrics (FA and RISH~2) while the differences in isotropy metrics (ADC and RISH~0) are smaller on average.
At this stage, it is unclear however if harmonization with the AIC regularization still contains variability attributable to the scanner
or if the CV criterion is too aggressive and mistakenly removed variability due to genuine anatomical variation.}
\reviewhbm{\cref{fig:percentage_diff_raw} shows the relative percentage difference that is originally present in the datasets between every pair of scanners,
but before applying harmonization and alterations.
This represents the amount of natural variability present in the diffusion metrics between scanners for each subjects in the region which is altered at a later stage.
We do not show the signal value for the \bval{0} and \bval{1200} images since the scanners are not using the same signal scaling.}

\begin{figure}
    \includegraphics[width=\linewidth]{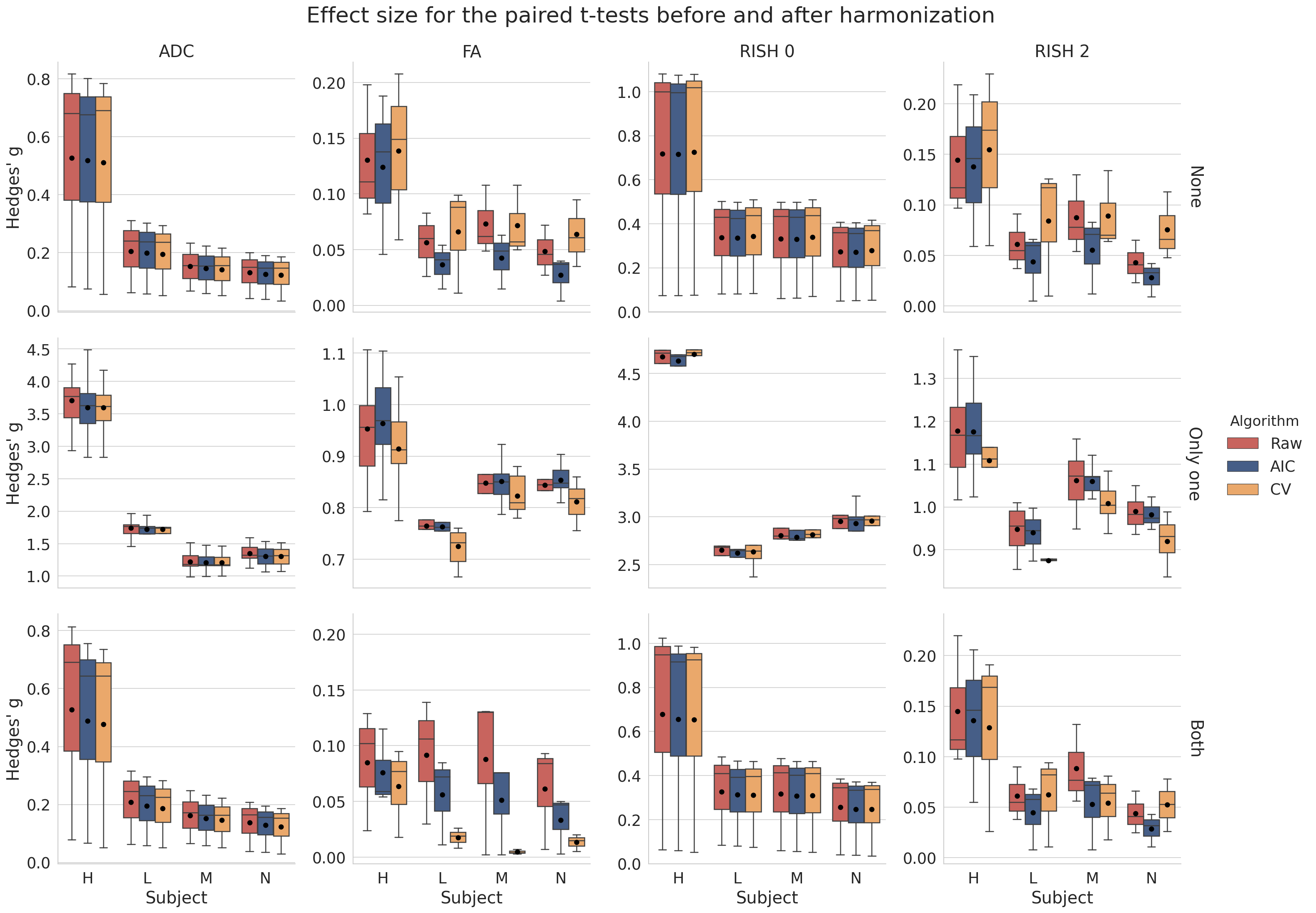}
    \caption{Boxplots of Hedges' g effect size for each metric with the mean value as the black dot.
    The raw data is shown in red (no harmonization), the data harmonized with the AIC in blue
    and finally the data harmonized with the CV in orange, similarly to the previous figures.
    The top row shows the effect size when both datasets are in their original version (\textbf{None} of the datasets are altered),
    the middle row when \textbf{only one} of the dataset is altered
    and the bottom row when \textbf{both} datasets are altered as indicated on the right of each row.
    The top and bottom row are only affected by scanner effects.
    The middle row shows larger effects size due to one of the compared dataset being altered in addition to the scanner effects.
    }
    \label{fig:effsize}
\end{figure}

\cref{fig:effsize} shows boxplots of the effect size as computed by a paired t-test after harmonization towards a common space for all scanners.
Tests were conducted for every subject between each scanner in addition to the altered versions of the datasets as previously described in \cref{sec:eval}.
For the ADC metric, both methods yield a lower effect size on average than the raw, unprocessed data
and preserve the effect size due to the alterations as shown in the middle row.
The RISH~0 metric shows similar behavior with the CV based method producing an average effect size slightly higher than the raw datasets.
Now looking at the anisotropy metrics (FA and RISH~2), the effect size is reduced or equal on average in most cases
(except for subject 'H' when only one scan is altered) when scans are harmonized with the AIC algorithm.
The CV based algorithm shows a higher effect size for harmonization between scans and a lower effect size when both scans are altered.
As we only report the absolute value of the effect size, this is due to both a lower group mean and group standard deviation
than the raw datasets.
\reviewhbm{This difference in group means and standard deviations prevents a direct comparison of results between each rows,
which can not be directly compared as they are unlikely to share a common numerator or denominator.}
The harmonization process is likely only removing scanner effects present in each dataset as the middle row
(where only one of the compared dataset is affected) shows similar reductions in effect size,
but is still on the same magnitude as the raw datasets since the alteration is preserved.

\begin{figure}
    \includegraphics[width=\linewidth]{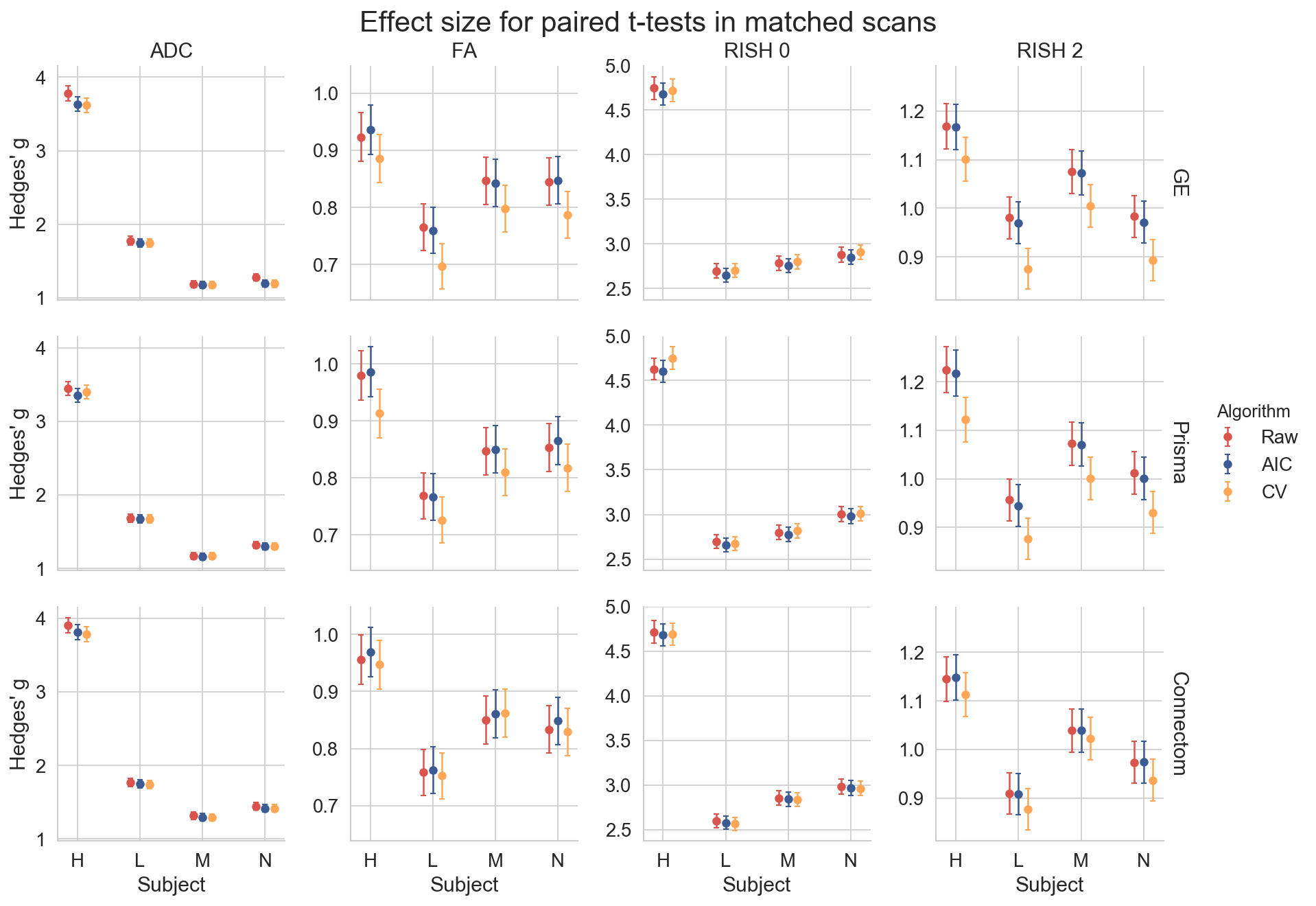}
    \caption{Hedges' g effect size for each metric between the original and altered datasets on the same scanner with a 95\% CI.
    The top row shows the effect size between the original and altered dataset on the GE scanner,
    the middle row for the Prisma scanner and the bottom row for the Connectom scanner.
    Most of the CI are overlapping except for the CV in the cases of subject 'L'
    on the GE scanner and subject 'H' on the Prisma scanner.
    This effect size is only due to the alterations performed in the experiments and is free of any other source of variability,
    such as registration error or scanner effects.
    }
    \label{fig:effsize_perscan}
\end{figure}

\cref{fig:effsize_perscan} shows the effect size, with a 95\% confidence interval (CI), for the paired t-test
between the original and altered datasets on each scanner.
While \cref{fig:effsize} showed the general trend for all results, we instead now focus on the effect size attributable solely to
the alterations we previously induced.
Results show that the ADC and RISH~0 metrics have the smallest CI, showing the lowest variability in the 3000 voxels in the altered region.
All CI are overlapping and therefore have a 95\% chance of containing the true mean effect size for every case.
The FA and RISH~2 metrics have both larger CI, showing larger variability in their sample values,
but are overlapping with the raw datasets CI in most cases.
Only the CV based harmonization method CI is outside the raw datasets CI for two cases.
This shows that the effect size is likely preserved after applying the harmonization algorithm in most cases
since the only source of variability is the effects we induced in that region to create the altered datasets.
The individual effect sizes, p-values and other intermediary statistics for every tested combination
that generated the boxplots shown in \cref{fig:effsize} \new{are available as Supplementary materials}.

\section{Discussion}
\label{sec:discussion}
\subsection{Reducing variability across scanners}

We have presented a new algorithm based on dictionary learning to harmonize datasets acquired on different scanners
using the benchmark database from the CDMRI 2017 harmonization challenge \citep{Tax2019}.
The flexibility of the method lies in its ability to \rereview{pool datasets from any scanner, without the need of paired datasets or spatial correspondence, by adapting}
the regularization parameter $\lambda_i$ automatically to each subset of training examples in \cref{eq:find_D},
ensuring that the relevant information to reconstruct the data is encoded in the dictionary $\bm{D}$.
Only features deemed important to the reconstruction are stored as the $\ell_1$ norm on the coefficients $\bm{\alpha}$ encourages a sparse reconstruction
and forces most of the coefficients to zero \citep{St-Jean2016a,Candes2008,Daubechies2010}.
In the reconstruction step, a new value of $\lambda_i$ is automatically selected for each reconstructed patch,
ensuring that the regularization is tuned uniquely so that the reconstruction matches the original patch, but using only features found in the target scanner.
This is of course at the cost of additional computations since a least-square problem needs to be solved for each candidate value $\lambda_i$,
but convex and efficient numerical routines reusing the previous solution as a starting point can be used to alleviate computational issues \citep{Friedman2010}.
To the best of our knowledge, this is the first case where an automatic search of the regularization parameter has been used in both stages of the optimization.

For the reconstruction step, we introduced two alternatives to compute $\lambda_i$ through the AIC or CV using held out parts of the signal.
While other choices are possible, such as the Bayesian information criterion \citep{Schwarz1978}, we chose here the AIC for simplicity and
because it is in fact equivalent to leave one out CV in the asymptotic limit \citep{Stone1977}.
Cross-validation was done with a classical approach as done in statistics, predicting the signal on parts of the current reconstructed patch as opposed to
\eg reconstructing a completely separate patch with the same value of $\lambda_i$ as may be done in machine learning.
This could explain why the AIC based method performed better than the CV criterion for the anisotropy metrics in the SA protocol since
the held out data, which is selected at random for every case, may sometimes unbalance the angular part of the signal because of the \review{random splitting process used during CV}.
The AIC would not be affected as it can access the whole data for prediction but instead penalizes reconstructions
that do not substantially reduce the mean $\ell_2$ error and are using too many coefficients---a likely situation of overfitting.
This also makes the AIC faster to compute since there is no need to refit the whole model from the beginning unlike the CV.
\rereview{While we used 3-fold cross-validation in this work to limit computations, better results may be obtained by increasing
the number of folds held out in total as additional data would be available at each step.
However, it is important to keep in mind that the whole model needs to be fitted K-times for K-fold cross-validation,
which may be prohibitive from a computational standpoint if many datasets are to be harmonized.}

One major advantage of the harmonization approach we presented is \reviewhbm{its} ability \new{to process raw datasets without the requirement of paired samples or spatial alignment during training.
In our experiments,} the data was given at random for the training phase and we mixed patches from all subjects and all scanners altogether in the additional experiments we described in \cref{sec:recon},
preventing overfitting to a particular scanner in the process.
\new{Other approaches instead go through an alternative representation such as the SH basis \citep{Mirzaalian2016a,CetinKarayumak2019,Blumberg2019}
or harmonize only the extracted scalar maps from diffusion MRI instead \citep{Alexander2017,Fortin2017a}.
In the latter cases}, it is not clear if the mapping developed for a particular scalar map is in fact similar between metrics as scanner effects may behave differently,
\eg isotropy metrics may be subject to global effects while anisotropy metrics may exhibit orientational bias due to low SNR in some given gradient directions.
We also observed in our experiments that the error for the ADC and RISH~0 metrics were similar for most methods
while the error was larger for the FA and RISH~2 metrics for the baseline method, which are orientation dependent.
This shows that the \enquote{optimal} mapping function could likely be task dependent if one wants to harmonize directly the scalar maps between scanners,
which could complicate interpretation between studies that are not using a matched number of b-values or gradient orientation.
\reviewhbm{In the original CDMRI challenge \citep{Tax2019}, the best performing algorithm for some cases of the anisotropy metrics was the baseline algorithm.
This was attributed to the blurring resulting of the SH basis interpolation in the angular domain with a trilinear interpolation when the spatial resolution of the datasets is not matching.
These results were obtained by applying the harmonization on the GE scanner datasets only while leaving the target scanners (Prisma and Connectom) datasets intact.
This task consists in matching the distribution from a source scanner to a target scanner, but without harmonizing the target scanner.
This blurring introduced by interpolation could also explain why the baseline method outperforms some of the compared algorithms for the KL divergence in \cref{fig:sa_kl,fig:st_kl}
as this SH interpolation step was not included in the AIC or CV algorithms of this manuscript.}

In the additional experiments, we introduced the idea of creating a neutral scanner-space instead of mapping the datasets towards a single target scanner.
We also harmonized datasets that had been altered towards that common space and \reviewhbm{showed} that the induced effect sizes are preserved
while at the same time preserving normal anatomical variability.
This approach has the benefit of removing variability attributable to \rereview{multiple} scanners,
instead of trying to force \rereview{a} source scanner to mimic variability that is solely attributable to \rereview{a} target scanner.
It is also important to mention here that a good harmonization method should remove unwanted variability due to instrumentation,
all the while preserving genuine anatomical effects as also pointed out previously by \citet{Fortin2017a}.
While this statement may seem obvious, success of harmonization towards a common space is much more difficult to quantify than \review{harmonization} between scanners
since we can not look at difference maps between harmonized datasets anymore.
As a thought experiment, a harmonization method that would map all datasets towards a constant value
would show no difference between the harmonized datasets themselves, therefore entirely removing all variability.
It would however commit very large errors when compared against the original version of those same datasets.
From \cref{fig:datagrid}, we see that the harmonized datasets are similar to their original version, but \review{\cref{fig:datagrid_diff,fig:percentage_diff}}
show that the CV based algorithm has larger relative differences with the data before harmonization.
It is however not obvious if the CV based algorithm is removing too much variability by underfitting the data
or if the AIC based method is not removing enough, overfitting the data.
\cref{fig:effsize_perscan} suggests that the CV criterion might underfit the data due to the lower effect size,
but this could be due to using only 3 fold CV in our experiments to limit computation time.
Results might be improved by using more folds as the AIC approximates the CV as we have previously mentioned.
\rereview{Our recommendation is therefore to use the AIC criterion on large cohort where computation resources are limited,
but improvements could be possible by increasing the number of folds for CV or even using a separate test set
to build the dictionary if enough data is available to do so.}

\subsection{Dependence of isotropy and anisotropy metrics on scanning parameters}
\label{sec:diffusion_dependence}

\begin{figure}
    \includegraphics[width=\textwidth]{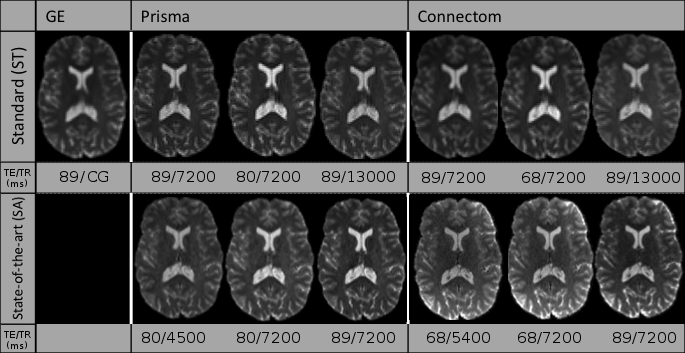}
    \caption{Example \bval{0} images for the standard protocol (top row) and the state-of-the-art protocol (bottom row)
    for a single subject acquired on the three scanners at different combinations of TE and TR.
    Note that the \bval{0} image for the GE scanner was only acquired at a single TE with a cardiac gated (CG) TR.
    The figure is adapted from \citet{Tax2019}, available under the CC-BY 4.0 license.}
    \label{fig:datasets_b0s}
\end{figure}

While it is usually advocated that protocols should use similar scanning parameters as much as possible to ensure reproducibility,
this is not always easily feasible depending on the sequences readily available from a given vendor and differences in their implementations.
Subtle changes in TE and TR influence the measured signal as shown in \cref{fig:datasets_b0s} by changing the relative T2 and T1 weighting of the measured diffusion signal, respectively.
While dMRI local models are usually applied on a per voxel basis, changes in these weightings will yield different values of the diffusion metrics,
which makes comparisons between scans more difficult as the weighting depends on the different (and unknown) values of T1 and T2 of each voxel \citep[Chap. 8]{Brown2014}.
Even if these changes are global for \reviewhbm{all} voxels, matched b-values are not sufficient to ensure that the diffusion time is identical between scans
as changes in TE influence diffusion metrics such as increased FA \citep{Qin2009}, but this effect may only manifest itself
at either long or very short diffusion times in the human brain \citep{Clark2001,Kim2005}.
Proper care should be taken to match the diffusion time beyond the well-known b-value, which may not always be the case if different sequences are used
\eg PGSE on the Siemens scanners and TRSE on the GE scanner as used in this manuscript.
Additional effects due to gradients and b-values spatial distortions \citep{Bammer2003} could also adversely affect the diffusion metrics,
especially on the Connectom scanner as it uses strong gradients of 300 mT/m \citep{Tax2019}.
Isotropy metrics are not necessarily free of these confounds as gradients nonlinearity create a spatially dependent error on the b-values \citep{Paquette2019}.
This could explain the larger mean error for the CV and baseline methods on the Connectom scanner harmonization task, especially for the anisotropy metrics.
While correcting for these effects is not straightforward, gradient timings should be reported in addition to the usual parameters
(\eg TE, TR, b-values and number of gradient directions) in studies to ease subsequent harmonization.
Accounting for these differences during analysis could be done \eg by using a (possibly mono-exponential) model
including diffusion time and predicting the diffusion metrics of interest at common scanning parameters values
between the acquisitions to harmonize.

\subsection{Limitations}
\label{sec:limitations}
\paragraph{Limitations of harmonization}

As \citet{Burnham2004} stated, \enquote{in a very important sense, we are not trying to model the data;
instead, we are trying to model the information in the data}.
This is indeed the approach taken in the challenge by the participants, the four other entries relying
on deep learning and neural networks for the most part with all methods (including ours) optimizing a loss function
which considered the difference between the original and the harmonized dataset.
With the rapid rise of the next generation of deep learning methods such as generative adversarial networks (GAN) and extensions \citep{Goodfellow2014},
it is now possible to instead model \rereview{implicitly} the \textit{distribution} of the data.
This allows generation of datasets from a completely different imaging modality such as synthesizing target CT datasets from source MRI datasets \citep{Wolterink2017}.
However, if proper care is not taken to \rereview{sample} truthfully the distribution of the data
(\eg not including enough tumor samples in a harmonization task between datasets with pathological data),
this can lead to severe issues.
\citet{Cohen2018} recently showed that in such a case, GAN based methods could try to remove the pathology in the data to match the distribution
of healthy subjects that the method previously learned, precluding potential applications to new datasets or pathological cases not represented \enquote{well enough}
in the training set.
The same concept would likely apply to systematic artifacts; if every dataset from a target scanner is corrupted by \eg a table vibration artifact,
it may very well be possible that a harmonization algorithm will try to imprint this artifact to the source datasets to match the target datasets.
The same remark would apply to our harmonization algorithm; if systematic artifacts are in the data, the learned dictionary may very well try to reconstruct
these systematic artifacts.
However, when rebuilding the source dataset using this corrupted target dictionary, we expect that the artifact would be mitigated since it would not appear in the
source dataset and hence should not be reconstructed by \cref{eq:find_D} as it would penalize the $\ell_2$ norm part of the cost function.
\rereview{This remark also applies to normal variability of the subjects; if the training datasets are too heterogeneous
(\eg young and healthy subjects mixed in with an older population affected by a neurological trait of interest),
harmonization algorithms may mistakenly identify (and subsequently remove) information attributable to biological differences between subjects rather than scanner variability.
It is therefore implicitly assumed in our algorithm that the datasets to harmonize are representative
and well matched (\eg age, gender) when removing scanner-only differences as other sources of expected variability
can be alternatively included in the statistical testing step of the study at hand.}
While offering a promising avenue, care must be taken when analyzing harmonization methods to ensure that they still faithfully represent the data
as optimal values of the cost functions themselves or \enquote{good} reconstruction of the diffusion metrics only
may not ensure this fact \citep{Rohlfing2012a}.

\paragraph{Limitations of our algorithm and possible improvements}

Our additional experiments with simulated free water have shown how harmonization can, to a certain extent, account for data abnormalities not part of the training set.
However, the presence of CSF and the boundary between gray matter and CSF (or a linear combination of those elements)
may yield enough information for the reconstruction to encode these features in the dictionary.
This can provide new elements that are not used for the reconstruction of normal white matter but may be useful for the altered data in the experiments.
It is not necessarily true that this property would also be valid for other neurological disorders such as tumors, especially if their features are not well represented
in the training data as we have mentioned previously in \cref{sec:limitations}.
Another aspect that we did not explicitly cover is multishell data \ie datasets acquired with multiple b-values,
which was in fact part of the following CDMRI challenge \citep{Ning2019}.
Nevertheless, our method can still be used on such datasets, but would not be aware of the relationship between DWIs beyond the angular domain.
Other approaches to build the dictionary could be used to inform the algorithm and potentially increase performance on such datasets
by explicitly modeling the spatial and angular relationship \citep{Schwab2018b} or
using an adaptive weighting considering the b-values in the angular domain \citep{Duits2019} amongst other possible strategies.
\review{This weighting strategy could be used for repeated acquisitions or if multishell datasets without an equal repartition of the data across shells needs to be harmonized
instead of the strictly angular criterion we used in this manuscript.
Note however that redefining the extraction step would only affect the initial creation of the patches as defined in \cref{sec:appendix}, leaving \cref{eq:find_D} unchanged.}
Modeling explicitly the angular part of the signal could also be used to sample new gradients directions directly,
an aspect we covered in the original CDMRI challenge by using the spherical harmonics basis \citep{Descoteaux2007b}.
Correction for the nature of the noise distribution could also be subsequently included as a processing step before harmonization
since reconstruction algorithms vary by scanner vendor \citep{Dietrich2008,St-jean2018a},
leading to differences between scans due to changes in the noise floor level \citep{Sakaie2018}.
Improvements could also potentially be achieved by considering the group structure shared by overlapping patches when optimizing \cref{eq:find_D} \citep{Simon2013a}.
While this structure would need to be explicitly specified, optimizing jointly groups of variables has recently led to massive improvements in other applications of diffusion MRI
such as reduction of false positive connections in tractography \citep{Schiavi2019}.
\rereview{In the end, the aim of harmonization procedures is to reduce variability arising from non-biological effects of interest in the application at hand.
Future benefits for this class of methods should therefore be evaluated on the end result of a study, rather than using proxy metrics of the diffusion signal for evaluation as is commonly done.
In the current work, registration errors or misalignment between subjects may influence negatively the evaluation of the algorithms as previously outlined in the CDMRI challenge \citep{Tax2019},
even though \textit{a priori} alignment is not an assumption of the presented harmonization algorithm.
Further validation of the proposed harmonization algorithm is therefore planned on a large-scale retrospective multicenter study
to evaluate the effect of harmonization on clinical outcomes.}

\section{Conclusions}
\label{sec:conclusion}

In this paper, we have developed and evaluated a new harmonization algorithm to reduce intra and inter scanner differences.
Using the public database from the CDMRI 2017 harmonization challenge, we have shown how a mapping
to reduce variability attributable to the scanning protocol can be constructed automatically
through dictionary learning using datasets acquired on different scanners.
These datasets do not require to be matched or spatially registered, making the algorithm applicable in retrospective multicenter studies.
The harmonization can also be done for different spatial resolutions through careful matching
of the ratio between the spatial patch size used to build the dictionary and the spatial resolution of the target scanner.
We also introduced the concept of mapping datasets towards an arbitrary scanner-space
and used the proposed algorithm to reconstruct altered versions of the test datasets corrupted by a free water compartment,
even if such data was not part of the training datasets.
Results have shown that the effect size due to alterations is preserved after harmonization, while removing variability attributable to scanner effects in the datasets.
We also provided recommendation when harmonizing protocols, such as reporting the gradient timings
to inform subsequent harmonization algorithms which could exploit these values across studies.
As perfect matching of scanner parameters is difficult to do in practice due to differences in vendor implementations,
an alternative approach could be to account for these differences through models of diffusion using these additional parameters.
Nevertheless, as the algorithm is freely available, this could help multicenter studies in pooling their data while removing
scanner specific confounds and increase statistical power in the process.

\FloatBarrier

\printbibliography

\clearpage
\appendix
\gdef\thesection{\Alph{section}} %
\makeatletter
\renewcommand\@seccntformat[1]{Appendix \csname the#1\endcsname.\hspace{0.5em}}
\makeatother

\section{The harmonization algorithm}
\label{sec:appendix}

This appendix outlines the harmonization algorithm in two separate parts.
\cref{alg:find_D} first shows how to build a target dictionary as depicted in the top part of \cref{fig:diagram}.
The bottom part of the diagram shows how to rebuild a dataset given the dictionary and is detailed in \cref{alg:rebuild_data}.
Our implementation is also freely available at \url{https://github.com/samuelstjean/harmonization} \citep{St-Jean2019e}.

\begin{algorithm}
    \SetAlgoLined
    \KwData{Datasets, patch size, angular neighbor}
    \KwResult{Dictionary $\bm{D}$}
    \vspace{0.2cm}
    \textbf{Step 1} : \emph{Extracting patches from all datasets}\;
        \ForEach{Datasets}{
            Find the closest angular neighbors\;
            Create a 4D block with a \bval{0} image and the angular neighbors\;
            Extract all 3D patches and store the result in an array $\Omega$\;
        }
    \textbf{Step 2} : \emph{Build the target dictionary}\;
        \While{Number of max iterations not reached}{
            Randomly pick patches from $\Omega$\;
            Solve \cref{eq:find_D} for $\bm{\alpha}$ with $\bm{D}$ fixed\;
            Solve \cref{eq:find_D} for $\bm{D}$ with $\bm{\alpha}$ fixed using \eg \citet[Algorithm 2]{Mairal2009b}\;
        }
    \caption{The proposed harmonization algorithm - building a target dictionary.}
    \label{alg:find_D}
\end{algorithm}

\begin{algorithm}
    \SetAlgoLined
    \KwData{Dataset, dictionary}
    \KwResult{Harmonized dataset}
    \vspace{0.2cm}
    \textbf{Step 1} : \emph{Extracting patches from the dataset to harmonize}\;
        \ForEach{Dataset}{
            Find the closest angular neighbors\;
            Create a 4D block with a \bval{0} image and the angular neighbors\;
            Extract all overlapping 3D patches and store the result as $\Omega$\;
        }
    \vspace{0.2cm}
    \eIf {Matching across spatial resolution}{
           Downsample $\bm{D}$ into $\bm{D_\text{small}}$ spatially before reconstruction­\;
        }{$\bm{D_\text{small}}$ = $\bm{D}$\;}
    \vspace{0.2cm}
    \textbf{Step 2} : \emph{Find the harmonized patch}\;
    \ForEach{patches $\in \Omega$}
    {
        Find the coefficients $\bm{\alpha}$ by solving \cref{eq:find_D} for $\bm{D}_\text{small}$ fixed\;

        Find the harmonized representation $\bm{X} = \bm{D \alpha}$\;
    }
    \ForEach{patches $\in \Omega$}
    {
        Put back each patch at its spatial location and average overlapping parts\;
    }
    \caption{The proposed harmonization algorithm - reconstruction of the harmonized data.}
    \label{alg:rebuild_data}
\end{algorithm}

\end{document}